\documentclass[sigconf]{acmart}

\usepackage{booktabs} 
\usepackage{caption}
\usepackage{subcaption}   
\usepackage{enumitem}

\setlist{nolistsep,leftmargin=*}

\renewcommand\footnotetextcopyrightpermission[1]{} 
\setcopyright{none}

\usepackage[printwatermark]{xwatermark} 
\usepackage{xcolor} 
\usepackage{graphicx} 
\usepackage{lipsum} 

\begin{document}
\title{Pointer-Chase Prefetcher for Linked Data Structures}

\author{Nitish Kumar Srivastava}
\affiliation{%
  \institution{Cornell University}
}
\email{nks45@cornell.edu}

\author{Akshay Dilip Navalakha}
\affiliation{%
  \institution{Cornell University}
}
\email{adn47@cornell.edu}

\begin{abstract}
Caches only exploit spatial and temporal locality in a set of address referenced in a program. Due to dynamic construction of linked data-structures, they are difficult to cache as the spatial locality between the nodes is highly dependent on the data layout. Prefetching can play an important role in improving the performance of linked data-structures. In this project, a pointer chase mechanism along with compiler hints is adopted to design a prefetcher for linked data-structures. The design is evaluated against the baseline of processor with cache in terms of performance, area and energy.
\end{abstract}

%
%



\maketitle

\section{Introduction}
\label{sec-introduction}

 In high-end processor systems, off chip memory access is a bottleneck. Caches are used to reduce the
number of memory accesses, but they are able to exploit only spatial and temporal locality among
the data. Various linked data structures such as linked list, trees and graphs do not have enough
spatial locality. These data structures are very common in various day-to-day applications. Most
of the programming languages provide implementations of various collections which are implemented
using linked lists and trees. Linked lists are also used for variety of purposes like creating stacks,
queues, graphs (Adjacency List), in hash table (each bucket of the table can itself be a linked list)
and in dynamic memory management. Various search engines, games, mobile-apps (such as chess,
FIFA) make ample use of linked data structures. Linked data structures provide flexibility of creating
and deleting nodes dynamically, and hence avoid the overhead of bothering about the size of the data
structure in advance. However, due to dynamic construction, they are difficult to cache as the spatial
locality between the nodes is highly dependent on the data layout. Caches only exploit temporal and
spatial locality in the set of addresses referenced by the program and hence, every first access to a
node for these type of these structures is generally a miss in the cache, which has a negative impact
on the performance.
Prefetching can play an important role in improving the performance of applications using linked
data structures. Various techniques make use of address stream regularities to extract arithmetic patterns
to make predictions for prefetching. However, these patterns are not necessarily found in linked
data access sequences. Much work has been done in the area of data prefetching, both in hardware
and software. Researchers have shown that compiler optimizations that improve data locality like
blocking and loop interchange can greatly reduce the need for prefetching \cite{mckinley1996improving}. However, they fundamentally
rely on compile time knowledge of the data-set layout and its interaction with the cache.
A low-cost hardware/software cooperative technique that enables bandwidth-efficient prefetching of
linked data structures has also been proposed \cite{ebrahimi2009techniques}. Research has been done to exploit the dependence
relationships between loads and stores, and prefetch the subsequent nodes on their basis \cite{roth1998dependence}. Here, a
dynamic scheme which captures the access patterns of linked data structures and also predict future
accesses with high accuracy has been adopted and the dependence relationships between loads that
produce addresses and loads that consume these address have been exploited. In \cite{hughes2001memory} the authors have studied memory-side prefetching technique to hide latency incurred
by inherently serial accesses to linked data structures (LDS). A programmable prefetch engine has
been used to traverse LDS independently from the processor. The prefetch engine is able to run ahead
of the processor to initiate data transfers earlier than the processor.

In this project, a mechanism to implement a prefetcher for linked-list type data structures is
explored. A pointer-chase mechanism along with compiler hints to prefetch the nodes of linked data
structure is adopted. The baseline design consists of a simple five stage pipelined processor designed
in ECE4750 interfaced with a blocking cache, which in turn is connected to memory. For alternate
design, a prefetcher is inserted between the cache and the memory system. When the processor receives
a pointer-based load instruction it will forward the read request to the cache. On a miss, the cache
will send the read request to the prefetcher. If it is a hit in the prefetcher, it will write the data
back into the cache in a single cycle and for a miss it will request the memory for the data at the
given address and will send the response back to the cache. At the same time, the prefetcher will
calculate the memory address of the next node and will make a read request to the memory. While
the read request for the next node is being processed by the memory, the processor will continue to
execute other instructions. On a normal load or store, the cache will send the read/write request to
the prefetcher which will forward the request to the memory but will not prefetch the next node. 

An incremental approach was adopted to implement the design. The datapath, control-path and the test
cases were developed in parallel to first implement normal read hit path followed by pointer chase
read path. After this the miss path for each of the request type was implemented. To test the prefetcher a novel testing strategy was adopted. Instead of following the traditional
way of testing for directed tests, the hit flag for each request was checked in the sink. By adopting this
testing strategy we ensured that the prefetcher was indeed prefetching the next node. This testing
strategy can be extended to any type of memory system like the cache lab in ECE4750.

To evaluate the effect of the prefetcher on the entire system a total of 9 micro-benchmarks were
used. Out of these 4 benchmarks used linked-data structures: linked-list traversal, linked-list insertion,
hashtable and hanoi. The alternate design performed better as compared to the baseline design for
linked-data structure benchmarks. Also, the percentage improvement in performance depends upon
the memory latency. As the memory latency increases the percentage improvement increases upto a
certain point and then decreases. The performance improvement around 30\% is observed over the
baseline design for most of the linked-list based micro benchmarks. For benchmarks having no linked
data-structures the performance degradation was less than 6\%. This degradation becomes negligible
with increasing memory latency. The area overhead of the prefetcher was estimated to be 7\%. The
energy consumption of the prefetcher was found to be around 60\% of the processor. This data does
not include the energy consumption in the cache and hence it is expected that the energy overhead
will be small in the entire processor-cache-prefetcher-memory system.

We thank Professor Christopher Batten for his continuous support and guidance throughout the
project. We would also like to thank Aashish Agarwal and Asha Ganeshan for sharing the area of the
synthesized cache which helped in our area evaluation.

\begin{figure}[h]
\centering
\includegraphics[scale=0.4]{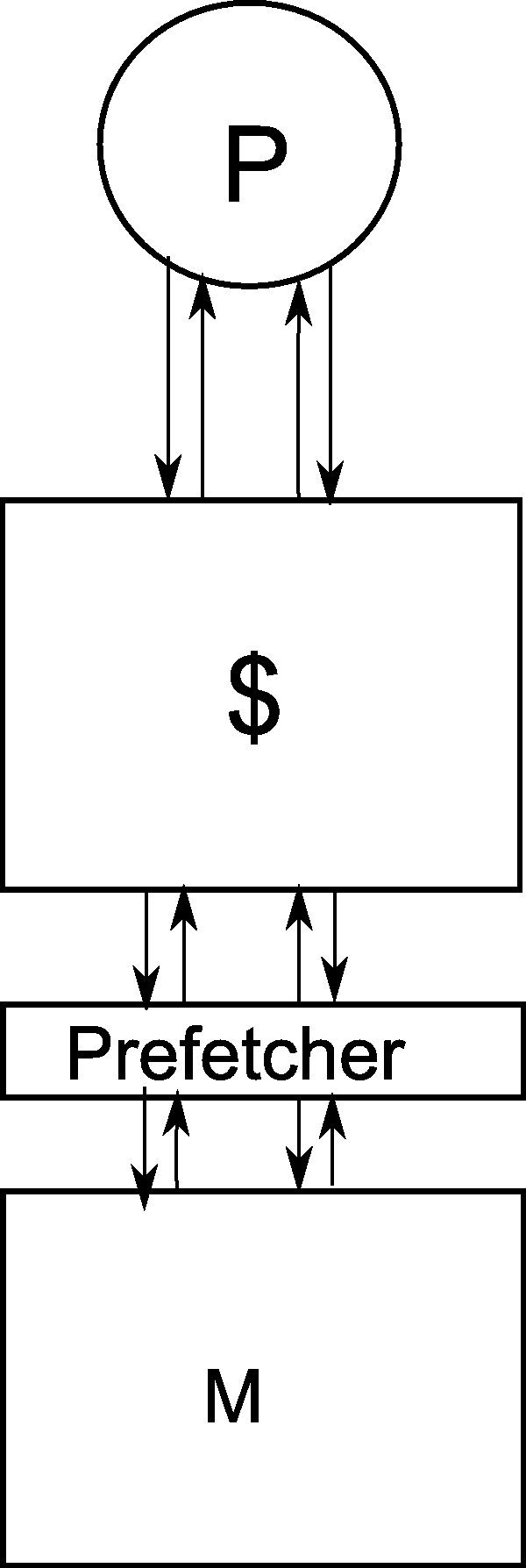}
\caption{Block diagram of Alternate Design}
\label{fig:block-diag-pfetch}
\end{figure}

The rest of the report is organized as follows:
Section~\ref{sec-baseline} discusses the baseline design with a processor and associated cache,
Section~\ref{sec-alternate} describes the proposed prefetcher design,
Section~\ref{sec-testing} discusses the testing strategy,
Section~\ref{sec-evaluation} discusses the evaluation;
followed by conclusions in Section~\ref{sec-conclusions}.

\section{Baseline Design}
\label{sec-baseline}

The baseline design for the project is a single core processor-cache-memory system. The block diagram
is shown in figure \ref{fig:block-diag}. It consists of a five-stage fully by-passed processor which is able to run all the instructions in the parc v2 ISA. The processor is connected to two direct mapped caches,
viz instruction and data caches. The instruction cache is connected to the instruction cache port of
the processor and the memreq0/memresp0 port of the test memory. The data cache is connected
to the data cache port of the processor and the memreq1/memresp1 port of the test memory. The
data bitwidth from processor to caches is 32 bits, while the data bitwidth from the caches to the test
memory is 128 bits (i.e., a full cache line). The message request and response format for different
modules is shown in the figure \ref{fig:mem-msg}. Each of the components are thoroughly tested using individual unit tests. A brief overview of each of the components are given below:

\begin{figure}[h]
\centering
\includegraphics[scale=0.3]{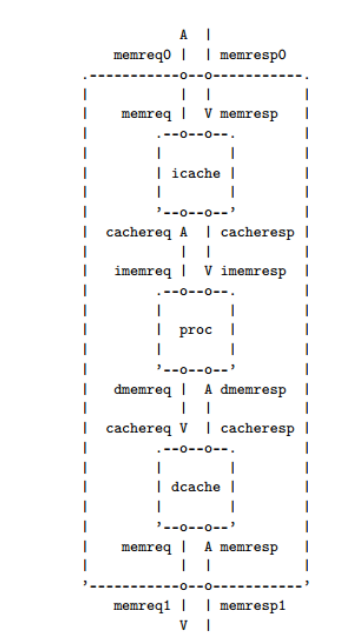}
\caption{Block diagram for Processor, Cache and Memory System}
\label{fig:block-diag}
\end{figure}

\begin{figure}[h]
\centering
\includegraphics[width=\columnwidth]{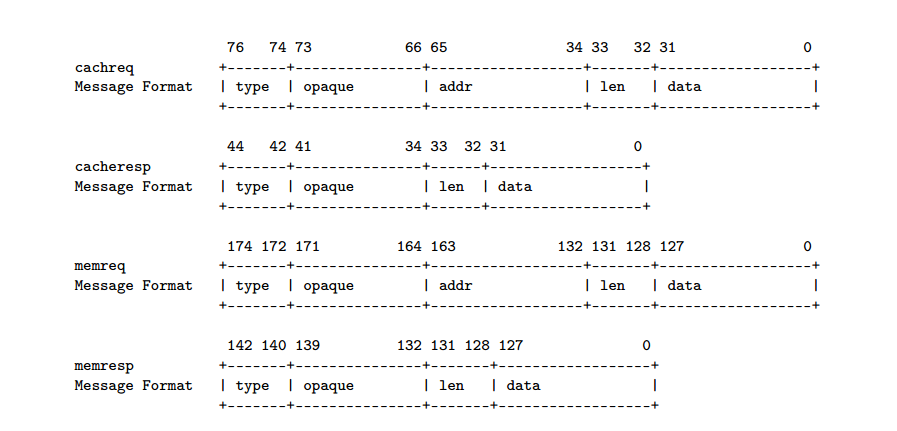}
\caption{Req/resp message format for Cache and memory}
\label{fig:mem-msg}
\end{figure}

\subsection{Processor}
\begin{figure}[h]
\centering
\includegraphics[width=\columnwidth]{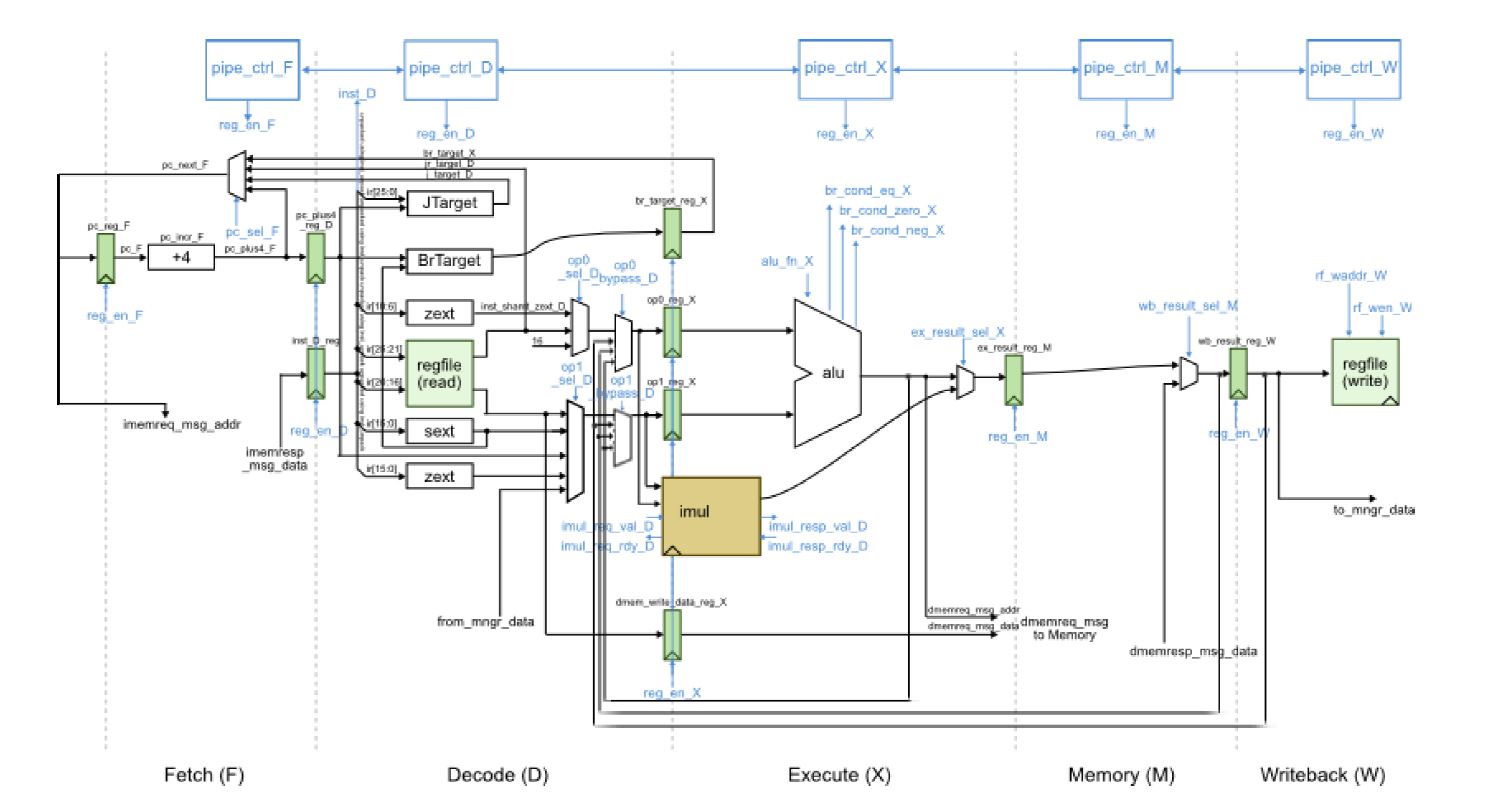}
\caption{Processor Datapath}
\label{fig:proc-datapath}
\end{figure}

A five stage pipelined processor with Fetch, Decode, Execute, Memory and Write
Back stages has been implemented. The Fetch stage fetches the instruction at the current PC from
the instruction memory, Decode stage mainly does two tasks i.e. decodes the instruction sent by the
Fetch stage and read the operand data from the register file. Apart from these two functions, Decode
stage also evaluates the jump and the branch targets for the jump and branch instructions. The
Execute stage does the ALU and multiplication operations and also checks for different branch taken
conditions. The Memory stage reads and writes the data from and into the data memory. The Write
Back stage writes the data to the register file. The datapath also consists of two bypass muxes, ALU
unit, a zero and a sign extension unit. The datapath diagram for the processor is shown in figure \ref{fig:proc-datapath}.
The control unit for the processor consists of the a control table for different signals depending upon
the instruction, the bypass and the stall logic to tackle RAW hazards.

\subsection{Cache}
\begin{figure}[h]
\centering
\includegraphics[width=\columnwidth]{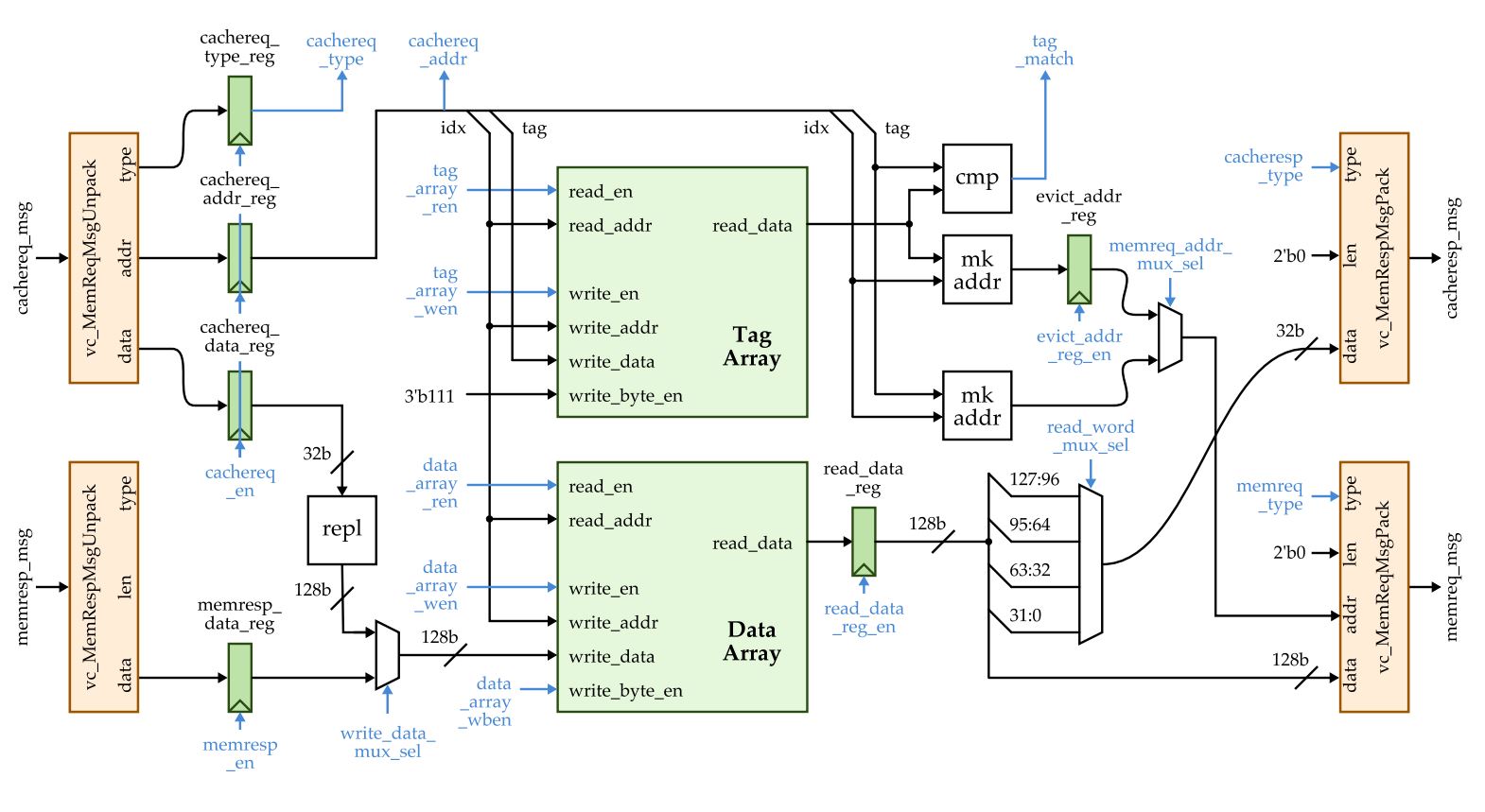}
\caption{Cache Datapath}
\label{fig:cache-datapath}
\end{figure}

\begin{figure}[h]
\centering
\includegraphics[width=\columnwidth]{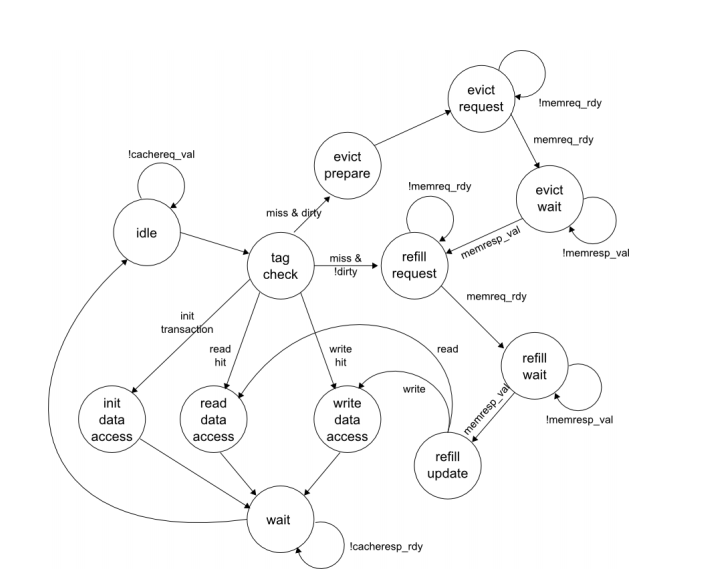}
\caption{Cache Control Unit}
\label{fig:cache-control}
\end{figure}

A direct mapped blocking-cache with a total capacity of 256 bytes and 16 bytes per cache
line has been implemented. The cache implements a write-back write allocate policy. The datapath
and control unit are shown in figure \ref{fig:cache-datapath} and figure \ref{fig:cache-control} respectively. The cache is thoroughly tested by directed and random testing. A parameterized design is used to achieve caches of different capacities and block size.

\subsection{Memory System}
A combinational memory was provided which returns the value of a memory
request within one cycle. Our alternate design aims to reduce the miss penalty of a pointer chase load
by using a prefetcher. To accurately predict the performance advantage, we need a better model for
the memory, so we decided to use a pipelined memory. The memory is able to accept the request as
long as the pipeline stages are not stalled and response from the memory is received by the system
after a fixed number of cycles. The pipelined memory gives the advantage that next memory request
does not need to stall while the previous memory request is being processed. At present an inelastic
pipelined memory is implemented, this can modified to implement an elastic pipeline to further improve
the throughput of the memory system. The pipelined stages are implemented as a separated
module outside the combinational memory to use hierarchy and abstraction principles. This design
methodology allows the designer to take advantages of the updates in individual components made
in the future.
A Val/rdy protocol is implemented to connect the processor and the cache, the cache and the
memory system, and the combinational memory and the pipeline stages, to make the interface latency
insensitive. The latency insensitive interface provides abstraction layer between the modules which makes the interface independent of the internal implementation of the different modules. This
allows to change the micro-architecture of the individual modules without affecting the functionality
of the entire system.

\section{Alternate Design}
\label{sec-alternate}

The alternate design is a modified version of the baseline consisting of a pointer chase prefetcher. The
prefetch module is inserted in between the cache and the memory system using a val/rdy interface.
In the baseline design, the processor can send two type of request to the memory system ( comprising
of cache and the memory ) viz. read and write type. In the alternate design, we have added a new
instruction load word chase pointer ( lw.cp ) with the following semantics:

\textbf{Semantics}

$lw.cp\,\,\,rd\,\,\,dest,\,\,\,i\,\,\,offset( r base )$ \\

$R[rdst]\,\,\,=\,\,\,M[R[rbase]\,\,\,+ \,\,\,sext(i offset) ]$

The semantics for the lw.cp is same as that of lw, with the difference that it provides a hint to the
hardware that the data read from the memory system contains the address of another node in the
linked data structure, which could be accessed in near future by the processor. Hence, at the same
time while doing lw.cp, the prefetcher will prefetch M [ M[ R[r dest] ] ]
This new instruction allows the processor to send a read chase pointer (read cp) request to the
memory system ( now comprising of the cache, prefetcher and the memory ). So now we have three
types of request: read, write and read cp

The processor will forward one of these three request to the cache. If it is a hit in the cache, it
will directly send the response back to the processor. On a miss, the cache will forward the request
to the prefetcher.
The prefetcher works in the following way depending upon the request type:

\begin{itemize}
    \item[1]\textbf{Write request from cache:} If it is a hit in the prefetcher, then invalidate the data in the prefetcher and forward the write request to the memory.
    \item[2]\textbf{Read request from the cache:} If it is a hit in the prefetch, the prefetcher sends the data to the cache. On a miss, it forwards the request to the memory and waits for its response.
    \item[3]\textbf{Read chase-pointer request:} On a hit, the prefetcher sends the data to the cache. It also determines the address of the next node and prefetches that data line from the memory. On a
miss in the prefetch, it forwards the request to the memory and waits for its response. When
the memory sends back the response, the prefetch forwards the response back to the cache and
sends another request back to the memory to search for the next node.
\end{itemize}

\subsection{Software Implementation}
For the benchmarks for performance evaluation, linked data structures need to be implemented. Since
the processor has no operating system running on it we cannot run applications which make use of
system calls. So we cannot use the malloc function in our applications which is used in C to dynamically
allocate new nodes from the memory heap to the linked structures. To tackle this problem, we
have implemented an array of nodes with random linkage among them i.e. a free list, which will be
used to allocate new nodes from the memory. To implement the random linkage among the nodes
in the array, a C based implementation of a pseudo-random number generator which does not makes
use of a system call was used. In this way the use of malloc function was avoided and the linked
data structure was implemented without envoking system calls.
For the evaluation purpose we have implemented applications which do insertion, linked list traversal
and hash table implementation. We have also borrowed linked-list based solution of the famous
Tower of Hanoi problem from Prof. Christopher Batten, which we will be using as one of our
benchmarks as well. All the benchmark applications have macros which allow us to share the same
code for native build on the normal gcc compiler, builds on the cross compiler for PARC and the ISA
simulator for the implemented processor. We are able to test these benchmarks both natively and on
the cross compiler as well as the ISA simulator.
In this project we have focused on the pointer-chasing mechanism where the pointer to the next
node in the data structure is the very first entry in the node. This reduced overhead of bothering
about the offset between the data elements and the next pointer. For this purpose, we had to make
sure that the first entry in the node struct definition should be a pointer to the next node.
We wanted to use the new instruction load cp in writing the benchmarks for the system. For this,
we used inline assembly, which allows the programmer to hard-code some assembly into a C program.
Using this feature, we were able to generate assembly with our new instruction.

\subsection{Hardware Implementation}
The hardware implementation includes modifying of the processor decode logic so that its supports
the new instruction load cp. It also requires modifying the memory request and response package in
the vc-mem-msgs to support the new type of request i.e read cp. The cache module is also modified
so that the cache sends the tag, index and the offset bits to the prefetcher. The offset bits are needed
by the prefetcher to locate the address of the next-node of the linked data-structure.
Apart from these changes we had to implement the entire prefetcher. The details of the prefetcher
are given below.

\subsection{Prefetcher Module}
The prefetch module like any cache module also has the data and the tag array with an additional
module to calculate the next address and a queue responsible for sending the request to the memory
for the next node. The implementation of the prefetch module can be divided into two submodules
which are connected in the top module. The first submodule is the datapath, which consist of the
paths for the data through various sub-modules like tag array, data array, address generation unit and
multiplexers. The second one is the control unit, which determines the next state and controls the
data flow through the datapath. The top level module consist of a connection between the datapath
and the control modules. In order to unit test the prefetch a test harness is created which connects
the prefetch module with the test source, test sink and test memory.
An incremental approach of the design has been adopted. We first implemented the logic in the
data and control path for the init transaction, followed by the hit path and then the miss path.

\subsection{Datapath}
\begin{figure}[h]
\centering
\includegraphics[width=\columnwidth]{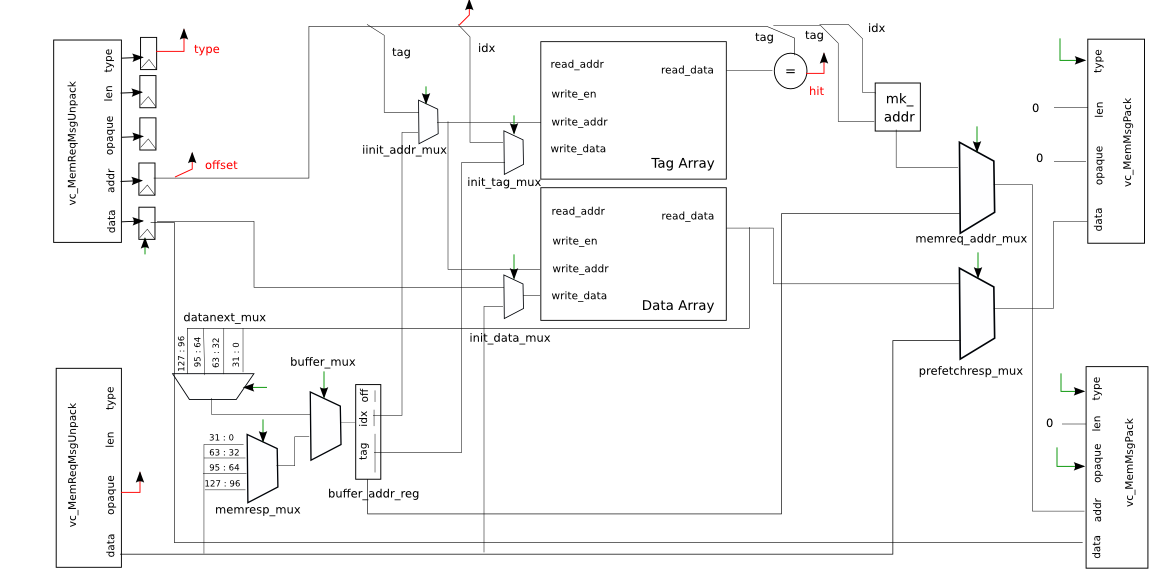}
\caption{Prefetcher Datapath}
\label{fig:prefetcher-datapath}
\end{figure}

The datapath for the prefetcher is shown in figure \ref{fig:prefetcher-datapath}. It can be broken into four parts i.e. message
unpack, message pack, next node address generation and the tag and data array.
The unpack module is responsible for extracting type, data, address and opaque fields from the
message packets. These values are then latched into a register. The tag and data array are used to store corresponding tag and prefetched data values in the
prefetcher. We have used registers for these arrays as their size is small ( total of 80 bytes in tag and
data arrays ) and so will not occupy a large area. The registers also have a faster response time as
compared to traditional SRAMs. Using the registers also provides an additional advantage of reading
and writing at the same time to these arrays and hence avoiding structural hazards ( as in SRAMs
using a single read/write port ). The tag-array checks if the incoming requested address matches with
the address in the tag-array. Depending upon a match, a hit signal is generated. On a hit in the
tag-array, data from the data-array is read and sent to the cache. On a miss, the mk-address module
is used to generate the address which is sent to the memory. The tag and the data-array contains
separate valid bits. When there is a read cp request the tag array is written with the address of the
next-node and the tag valid bit is set to one. The address generation module sends the corresponding
memory request for this tag. Till the response from the memory is obtained the data valid bit at
that corresponding index is set to zero. This design was used to ensure that if the prefetcher receives
read or read cp request which is already requested to the memory but is still in the memory pipeline,
the prefetcher will not send another request to the memory as the request already in the pipeline will
come back before. This would help to further boost the performance.
The address generation unit in the prefetcher consist of inputs from the memory and the data-array.
The unit consists of 3 multiplexers. Two of the multiplexers are used to extract the value of
the address of the next node from the corresponding line in the data array or the memory response,
depending upon the offset of the current read request. The third multiplexer is used to select the
value from the outputs of the two multiplexers. If the incoming request is of the type read cp and it
is a hit in the tag array the value from the data array is loaded in the buffer address register. When
there is a miss on a read cp request from the cache and read request is sent to the memory and the
prefetcher waits for the memory to respond. The corresponding address of the next node of the linked
data structure is determined by the address generation unit depending upon the offset sent by the
cache. The value of the next-node address is then latched in the buffer address register. In the next
cycle, this address is sent to the pipelined memory by the prefetch module and prefetcher becomes
ready to take the next request without waiting for the response from the memory. The buffer address
register holds this next-node address till the response from the memory is received. This helps to
write received data at the appropriate index in the data array.
As the memory is pipelined the prefetcher must keep a track of which responses it should send to
the address generation unit and write the data array ( i.e. the prefetched one ) and which responses
are to be sent back to the cache ( i.e. response to a cache request ). To track this the prefetcher
sets the opaque field appropriately while sending request to the memory. The opaque field is set to 1
whenever the prefetch address generation unit sends the request to the memory on a cache prefetch
miss and is set to be 0 when it is prefetching.

\subsection{Control Unit}

\begin{figure}[h]
\centering
\includegraphics[width=\columnwidth]{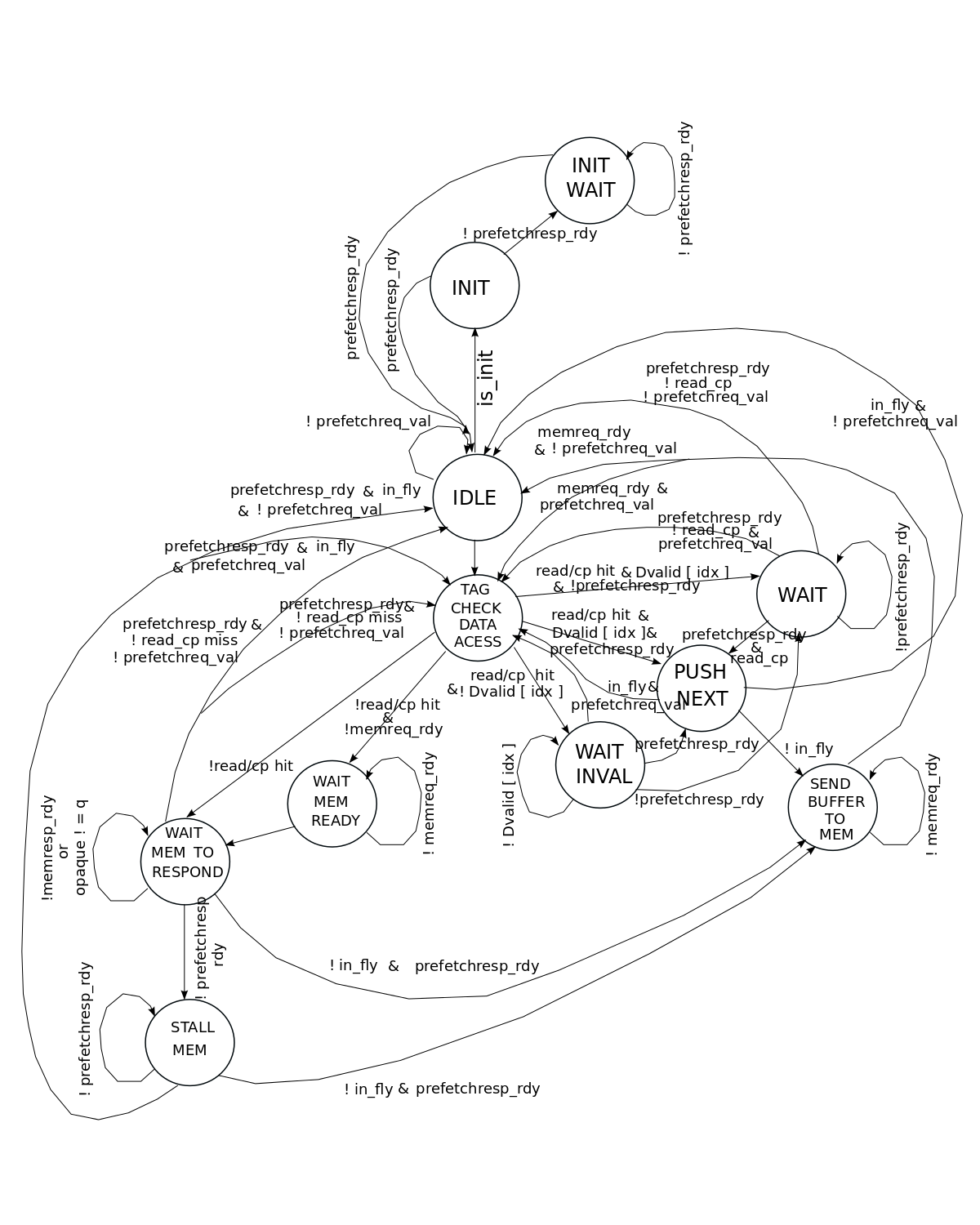}
\caption{Prefetcher Control Unit}
\label{fig:prefetcher-control}
\end{figure}

The control unit as shown in figure \ref{fig:prefetcher-control} uses an FSM to determine the state transitions and the control
signals for the datapath. The control path is mainly divided into 3 paths: The read/read cp miss and
write path, the read/read cp hit path and the init transaction path. A detail description of each of
the states is given below:\\

\begin{itemize}
    \item[1] \textbf{State Idle:} In this state the prefetcher is ready to accept the request from the test source by setting the prefetch req-rdy signal high. In this state, the registers are also enabled so that the incoming message can be latched whenever the req-val signal is set by the test source.
    \item[2] \textbf{State Tag-Check and Data-access:} In this state the tag from the request message is matched in the tag array and a hit signal is generated. Depending upon the request type and hit signal the control path diverges into different stages.
    \begin{itemize}
        \item[a] \textbf{Init Request:} The control path will move to the Init state.
        \item[b] \textbf{Read Hit:} If prefetcher response is ready and the valid bit of the data array is set, the data to the sink is sent in this state. This will ensure a single cycle hit latency. The prefetch req-rdy signal would be set high to accept a new request the next cycle. If data-array valid bit is low or the prefetch response is not ready the control flow will move into wait invalid or wait state respectively.
        \item[c] \textbf{Read CP-Hit:} This will follow a similar control signals as the read-hit. But in the next cycle it will move to the push-next state. In this case the prefetch would not be ready to accept a new request.
        \item[d] \textbf{Read-miss, read cp-miss or write:} If memory req-rdy is high it will move into the wait state till memory response comes back.
    \end{itemize}
    \item[3] \textbf{State Push-next:} In this state, the control signals of the address generation unit of the prefetcher are set. Depending upon hit and the offset of the read cp address, data from the data array or memory response is loaded in the buffer address register. The enable of the buffer address register is set high if there is no pending next-node prefetch. If there is a next-node address in the pipelined memory the buffer address register will not be able to accept the new pointer address and is dropped.
    \item[4] \textbf{State Buffer to Memory:} In this state the address in the buffer address register combined with an opaque bit set to 1 is sent to the memory to fetch the next-node. The buffer register address enable would be set to low to hold the address till its response is received from memory. Also the prefetch req-rdy is set high so that the prefetch can accept a new request from the source in the next cycle. If there is a prefetch request the next cycle the control flow will move into the tag-check state or else into the idle state.
    \item[5] \textbf{State Wait for memory to respond:} On a read miss, read cp miss or a write hit/miss the control flow will move to this state if the memory was ready to accept the request. The control flow will remain in this state till the memory response is valid and the opaque field is zero. This allow us to differentiate between memory request which comes from the address generation unit of the prefetcher and other requests. If the prefetch response is not ready the prefetcher will move into the stall memory state. The other scenario would be when the prefetch response is ready and the prefetch request is of type read cp and there is no buffer address register in the pipelined memory the prefetcher will go to the push-next state. For all other cases the prefetcher will go to the idle or tag-check state depending on whether there is a prefetch request in the next cycle or not.
    \item[6] \textbf{Stall memory:} The prefetcher will remain in this state till the prefetch resp-rdy becomes high i.e. the sink is able to accept the prefetch response. This state is necessary as if the memory is not stalled in this state, the memory response will be lost. On the similar condition as the wait states the prefetcher will move to the push-next state or tag-check state or idle state. 
\end{itemize}

\begin{figure}[h]
\centering
\includegraphics[width=\columnwidth]{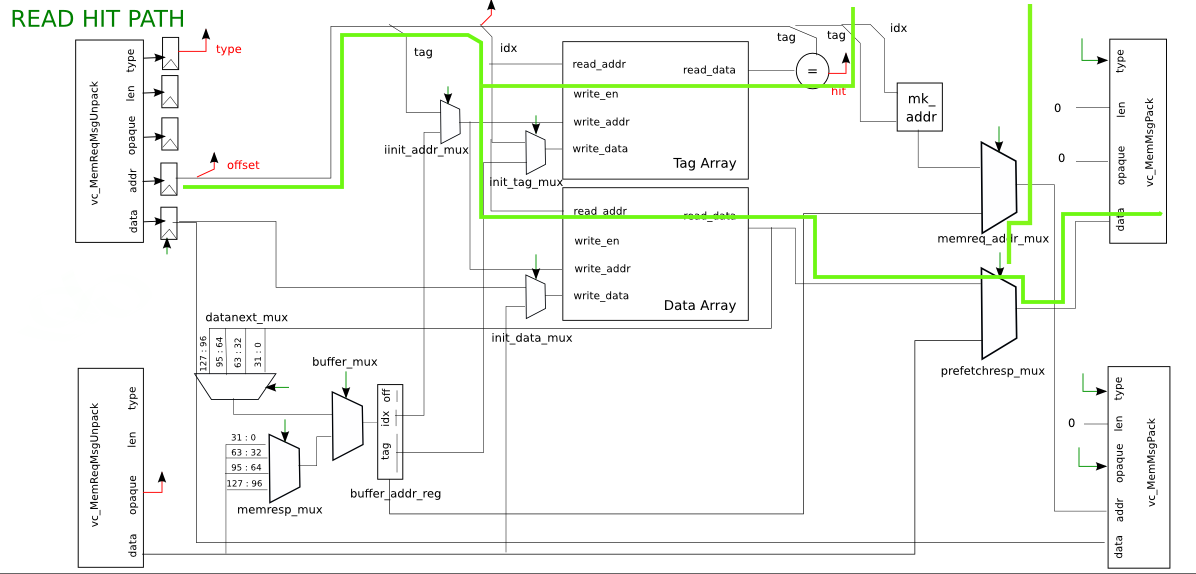}
\caption{Datapath showing the data and control flow for Read Hit}
\label{fig:read-hit}
\end{figure}

\begin{figure}[h]
\centering
\includegraphics[width=\columnwidth]{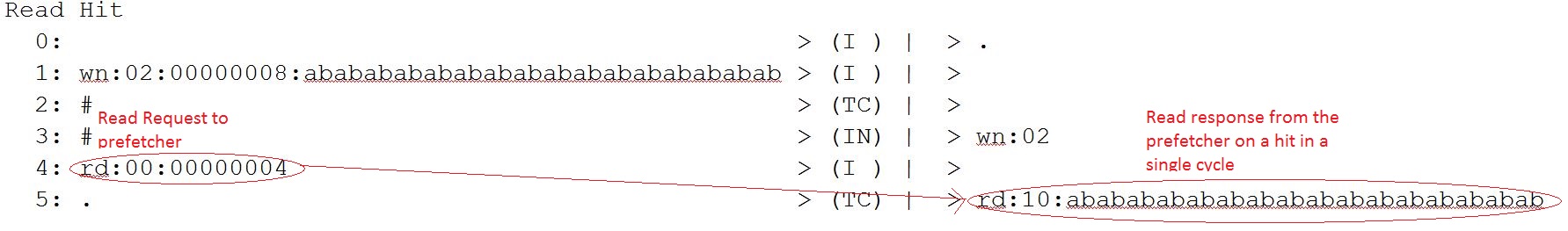}
\caption{Trace for Read Hit request}
\label{fig:trace-read-hit}
\end{figure}

\begin{figure}[h]
\centering
\includegraphics[width=\columnwidth]{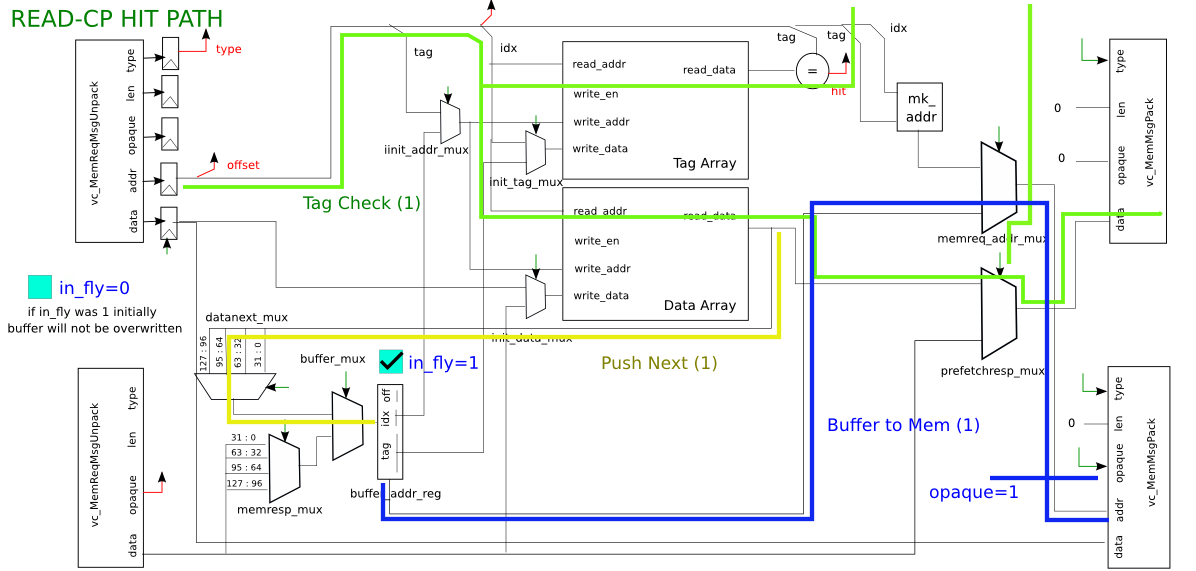}
\caption{Datapath showing the data and control flow for Read CP Hit}
\label{fig:read-cp-hit}
\end{figure}

\begin{figure}[h]
\centering
\includegraphics[width=\columnwidth]{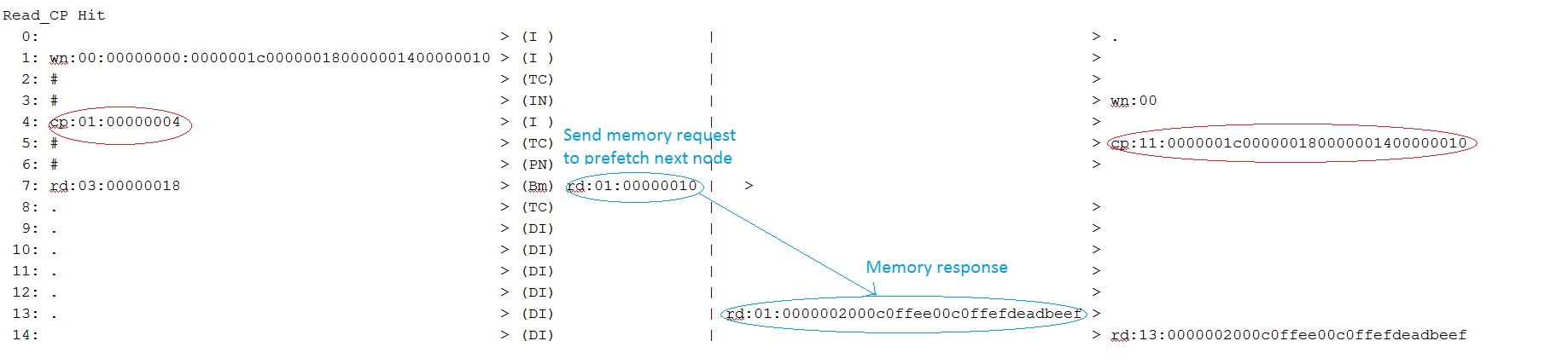}
\caption{Trace for Read CP Hit request}
\label{fig:trace-read-cp-hit}
\end{figure}

\begin{figure}[h]
\centering
\includegraphics[width=\columnwidth]{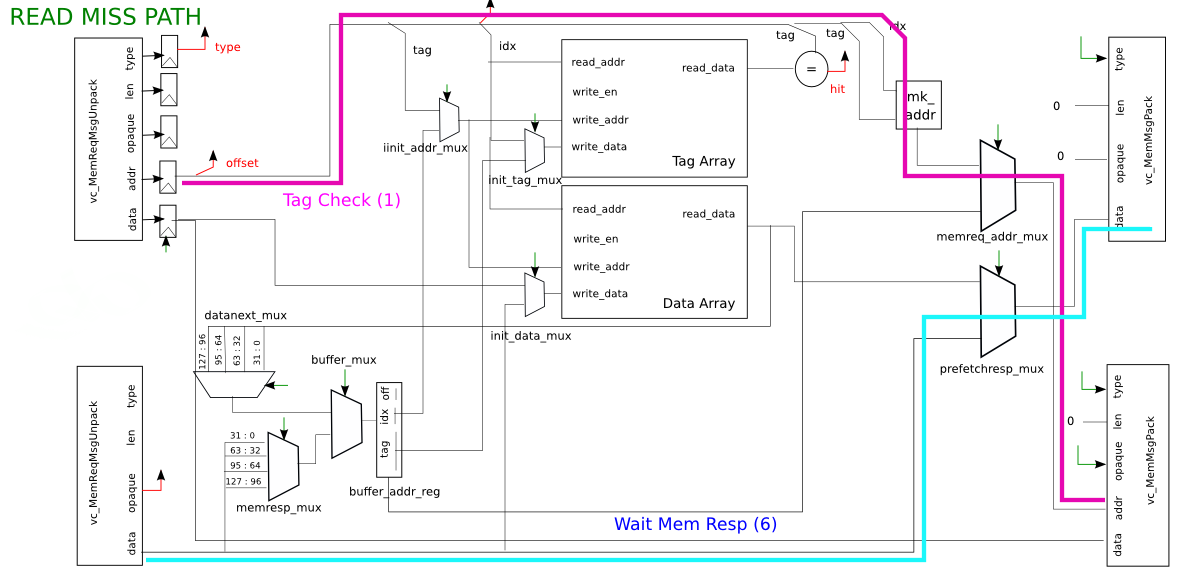}
\caption{Datapath showing the data and control flow for Read Miss}
\label{fig:read-miss}
\end{figure}

\begin{figure}[h]
\centering
\includegraphics[width=\columnwidth]{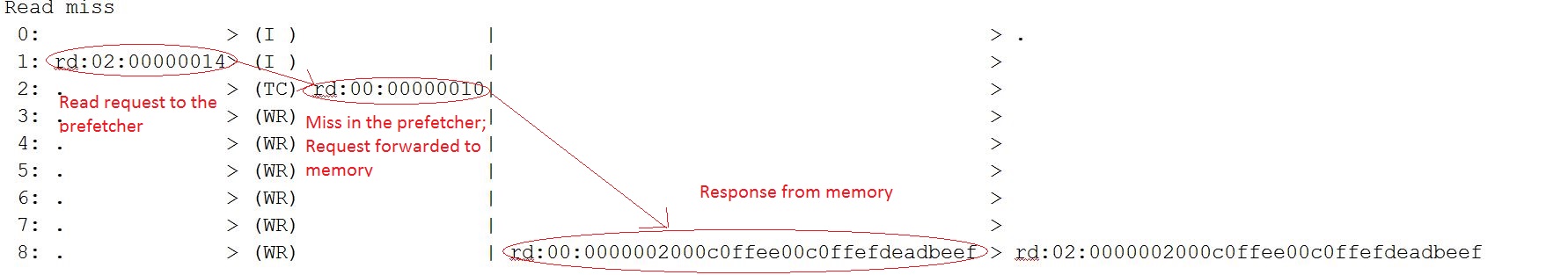}
\caption{Trace for Read Miss request}
\label{fig:trace-read-miss}
\end{figure}

\begin{figure}[h]
\centering
\includegraphics[width=\columnwidth]{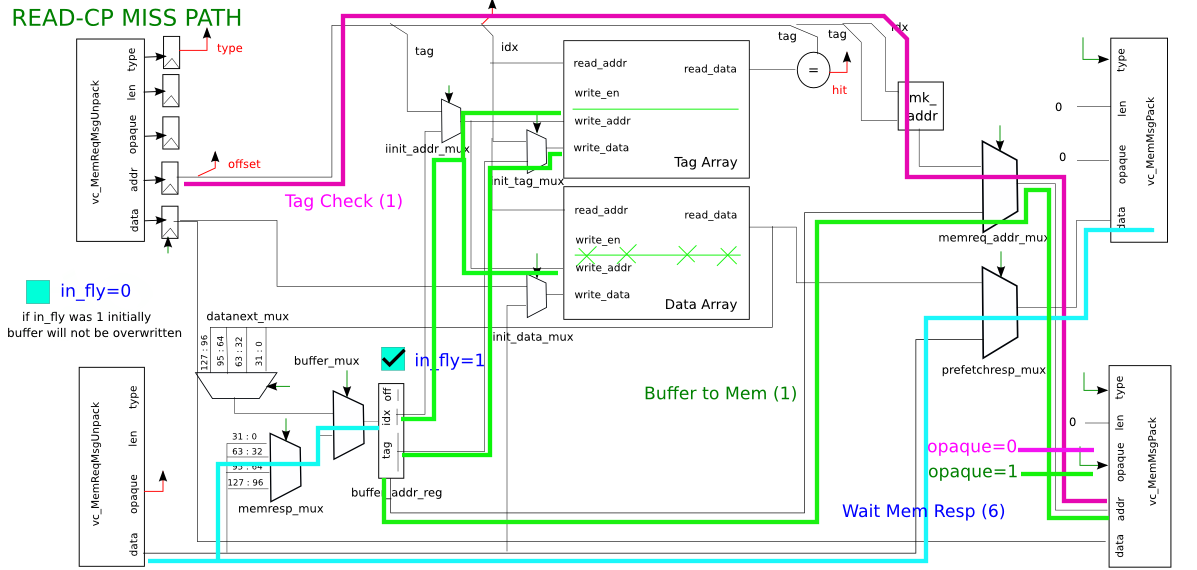}
\caption{Datapath showing the data and control flow for Read CP Miss}
\label{fig:read-cp-miss}
\end{figure}

\begin{figure}[h]
\centering
\includegraphics[width=\columnwidth]{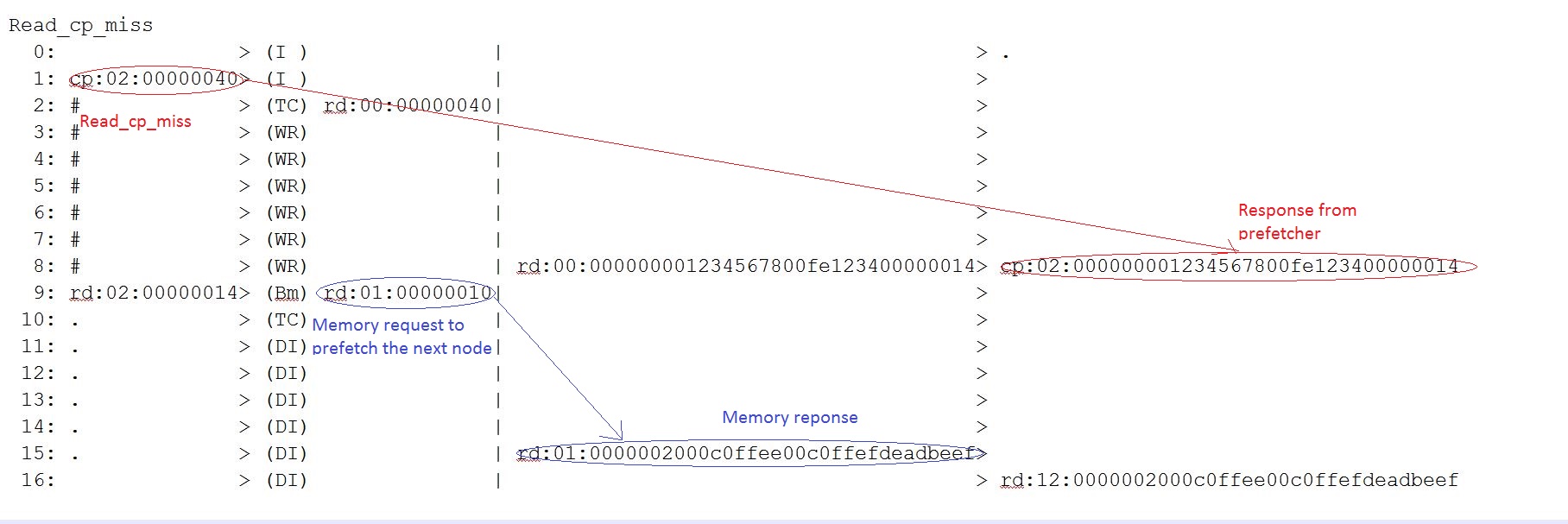}
\caption{Trace for Read CP Miss request}
\label{fig:trace-read-cp-miss}
\end{figure}

\begin{figure}[h]
\centering
\includegraphics[width=\columnwidth]{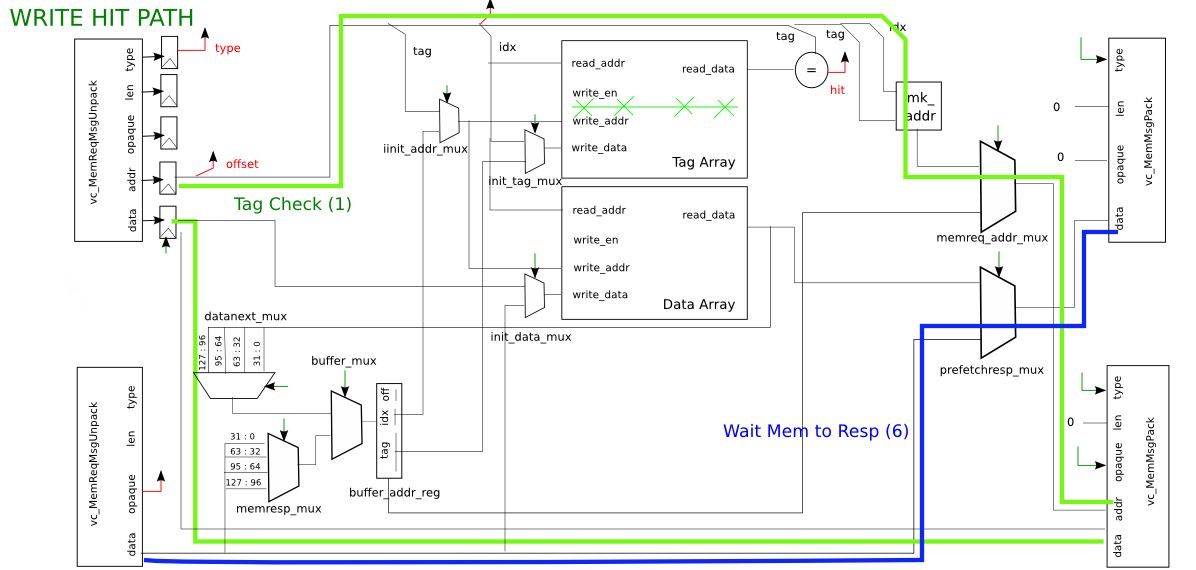}
\caption{Datapath showing the data and control flow for Write Hit}
\label{fig:write-hit}
\end{figure}

\begin{figure}[h]
\centering
\includegraphics[width=\columnwidth]{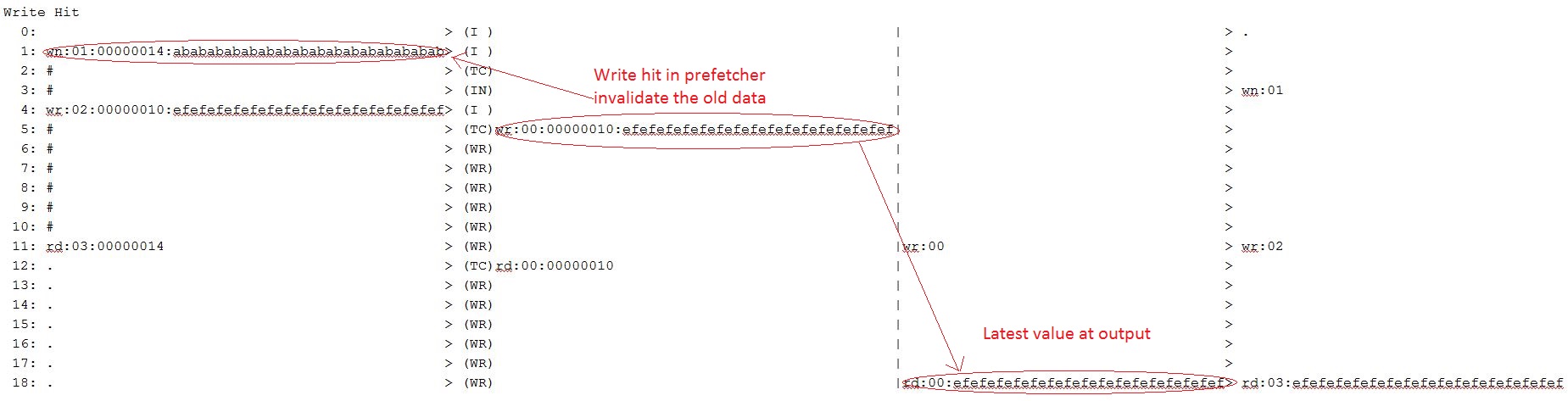}
\caption{Trace for Write Hit request}
\label{fig:trace-write-hit}
\end{figure}

\begin{figure}[h]
\centering
\includegraphics[width=\columnwidth]{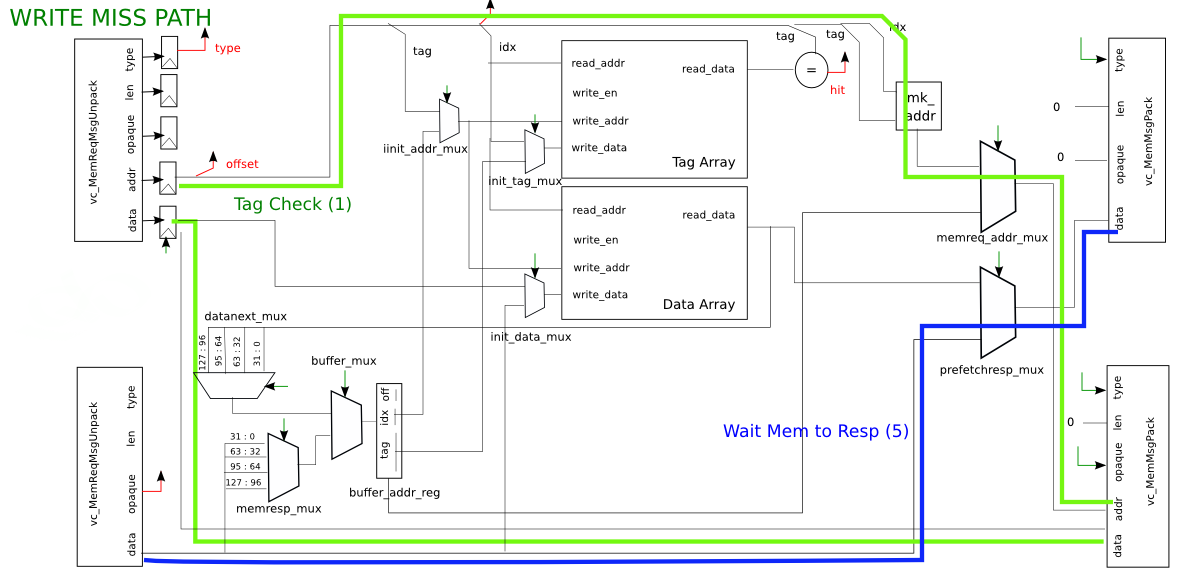}
\caption{Datapath showing the data and control flow for Write Miss}
\label{fig:write-miss}
\end{figure}

\begin{figure}[h]
\centering
\includegraphics[width=\columnwidth]{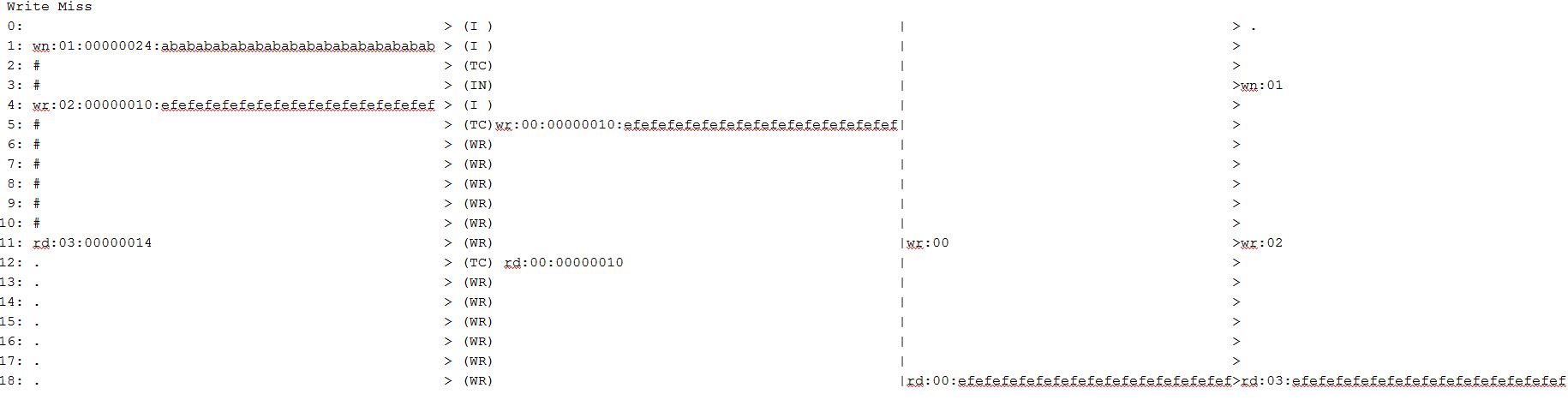}
\caption{Trace for Write Miss request}
\label{fig:trace-write-miss}
\end{figure}

The datapath diagram showing the data flow for each of the six possible cases: Read Hit, Read CP Hit, Read Miss, Read CP Miss, Write Hit and Write Miss are shown in figure \ref{fig:read-hit}-\ref{fig:trace-write-miss}. The different states are marked by different colors and the name of the states are also provided. Each datapath diagram is followed by a trace corresponding to that request. From the trace in figure \ref{fig:trace-read-hit} it can be seen that the read hit request can be handled by the prefetcher in single cycle which leads to single cycle hit latency in the prefetcher. 

    From the trace in figure \ref{fig:trace-read-cp-hit}, it can be seen that when a cp request is a hit in the prefetcher (circled in red), a prefetch request is sent to the memory for the next node (circled in blue). The trace also shows the critical case where the next request from the cache to the prefetcher is the read request to this next node. As it can be seen from the trace the prefetcher is stalled in the DI (Data Invalid) state waiting for the response of the prefetch request from the memory. The trace in figure \ref{fig:trace-read-miss} shows a read request from the cache to the prefetcher which is a miss. As it can be seen from the trace that prefetcher leads to an overhead of 1 cycle in case of a miss in the TC state.

    The trace in figure \ref{fig:trace-read-cp-miss} shows a read cp request which is a miss in the prefetcher (circled in red). As it can be seen from the trace that when the memory response for the cp miss arrives, the prefetcher sends another read request to the memory for the next node.
    
    Figure \ref{fig:trace-write-hit} shows the trace for the Write Hit path. The first request to the prefetcher is an init transaction. The second request is a Write Hit which invalidates the data in the corresponding line. This can be verified by the prefetch response data for read request at the same address.
    
\begin{figure}[h]
\centering
\includegraphics[width=\columnwidth]{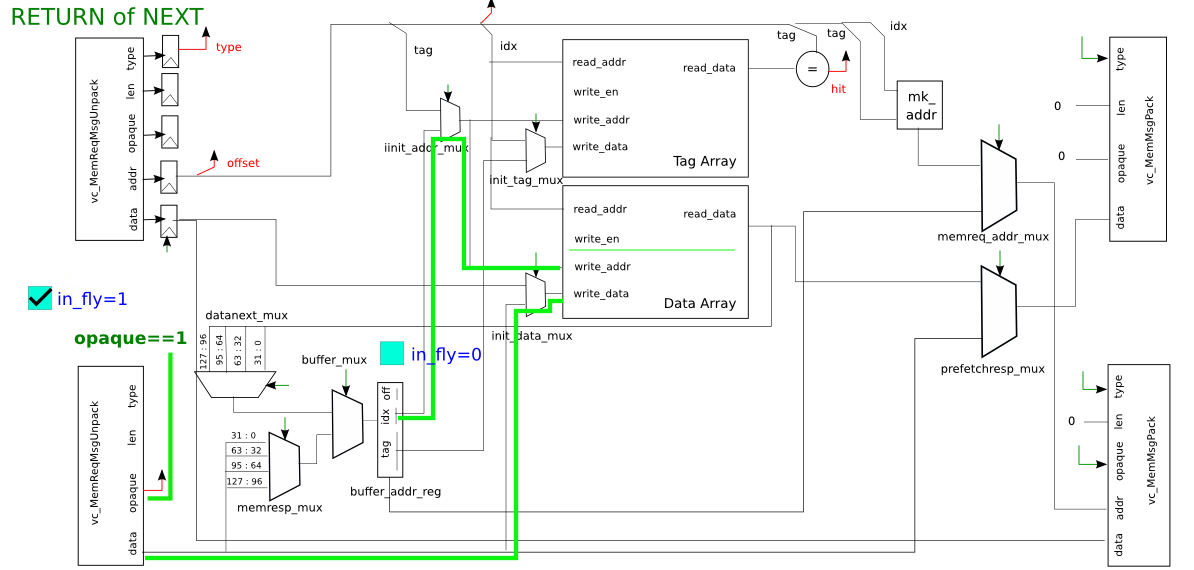}
\caption{Datapath showing the data and control flow for Handling the memory response for the
prefetched data (next node)}
\label{fig:return-next}
\end{figure}

\section{Testing Strategy}
\label{sec-testing}

The system is tested using two strategies viz. source-sink based testing and self-checking assembly tests. In source/sink based testing,
the values and the particular instruction to be tested are provided to the processor from the manager
and the result is verified in the sink. In assembly based testing, the instructions are encoded in
PARCv2 ISA format and are directly written in the instruction cache. The processor then fetches
the instructions from the cache in sequence and executes them. Upon execution, the processor makes
a jump to a pass or fail function where the processor sends either a pass or fail signal to the sink.
The sink depending on this signal displays the result of the program execution to the user. These
self-checking assembly tests also check that the fetch stage of processor along with the instruction
cache interface works correctly. Same tests are used for both of these approaches of testing. The test
framework converts the tests written into assembly sequence as is needed for the self-assembly testing.
Before integrating the prefetcher in the system it is necessary to ensure that the prefetcher bug
free and functionally correct and is prefetching the next node whenever it receives a read-pointer
chase instruction. As with the testing of any memory type of system, the testing of the prefetcher
was tricky. This was because the system would still pass all the tests even when there is no prefetch
happening. To tackle this problem, the hit flag for each prefetch request was checked in the test-sink.
The test-sinked checked whether for each prefetch request if the expected outcome of the hit flag
matched the actual hit flag sent by the prefetcher. In this way, it was ensured that the tests would
pass only if the prefetcher was behaving as per specification i.e. prefetched the next node on a read
chase pointer request.
A combination of directed tests and random tests were used to ensure full functional correctness
and cover all the possible corner cases.

\subsection{Directed Testing}
Directed tests provide us with an incremental approach to debug our design. First of all tests corresponding
to testing of the basic prefetch paths were implemented. This included testing the read hit
path, read pointer chase hit path, read miss path, read chase pointer miss path, the write hit path
and write miss path. The chase pointer read hit and miss path were tested to ensure that the next
node is prefetched from memory. The write hit path was tested to ensure that if the cache is going
to update the memory it should invalidate the tag in the prefetcher. By doing this, coherence issue is
solved as the prefetcher will always return the most recent value to the cache.
After ensuring that the basic paths were working, tests were written to ensure that the interaction
between the FSM and parallel control logic which receives the prefetch response from the memory
while the prefetch is still receiving request for the cache/ test-sink is working correctly. The testing
in this part included if the prefetcher is able to handle both the response from the memory and the
cache request in parallel. The cases that were considered were:

\begin{itemize}
    \item[1] The test-source request the data from an address whose memory request has been forwarded by the pefetcher in the previous cycles but has not receive the response from the memory. So in this case, the request would be a hit in the prefetcher but the data would invalid as the prefetcher is still waiting for memory response.
    \item[2] Another case would be that the prefetch receives the memory response in the same cycle the test-source sent the request for that node. This case would ensure correct data is sent at the output.
    \item[3] All the above tests were done with random source and sink delays to take care of the case where the memory would stall because the test-sink is not ready to accept the prefetch response which is forwarded from the memory.
\end{itemize}

Apart from these directed tests, long test of node traversal were written to see how well the prefetcher
responds to continuous streams of inputs as it would experience when integrated in the system.

\subsection{Random Testing}
Random testing is used to ensure nearly 100 percent functional coverage. It helps to cover all those
test cases which might be left out by directed tests. Random tests with different percentage of read,
read chase pointer and write requests were written. As the tests were random we do not know if the
given request would be hit or a miss. So to incorporate random testing tick-define in verilog was used
which ensured that the hit flag is not compared when we used random tests. Thus the random test
were used to ensure functional coverage of the tests. Thus by using directed and random tests the prefetcher was ensured to be bug free.

\section{Evaluation}
\label{sec-evaluation}

The baseline and the alternate designs are evaluated in terms of performance over a variety of microbenchmarks
having different percentage of pointer chase load instructions.
First, the benchmarks provided by the instructor were executed on both the baseline and alternate
design. These benchmarks had no linked data structures in them. In these benchmarks, the
performance of the alternate design is slightly degraded as compared to the baseline. The performance
difference between the alternate and baseline design depends upon the number of memory
operations. As the number of memory operations in the benchmarks increase the performance gap
between the alternate design and the baseline increases. As there are no linked data-structures used
in these programs, the prefetcher does not perform any useful work. On the other hand, it increases
the miss-penalty on a cache-miss by a single cycle. The prefetcher also affects the evict time of the
cache line by a single cycle.
As the memory latency increases, this overhead of 1 cycle seems to have a negligible effect on the
performance of the alternate over the baseline design. This variation can be seen from the figure \ref{fig:perf-degrade}.
It can be observed that as memory latency increases from 2 to 40 cycles the performance degradation
of the alternate design over the baseline design reduces. The degradation is less than 2\% for memory
latency of 40 cycles on all the benchmarks. In real world, the memory latency is of tens of cycles.
So, we expect that inserting the prefetcher in the memory hierarchy of the system will not affect its
performance significantly.

\begin{figure}[h]
\centering
\includegraphics[width=\columnwidth]{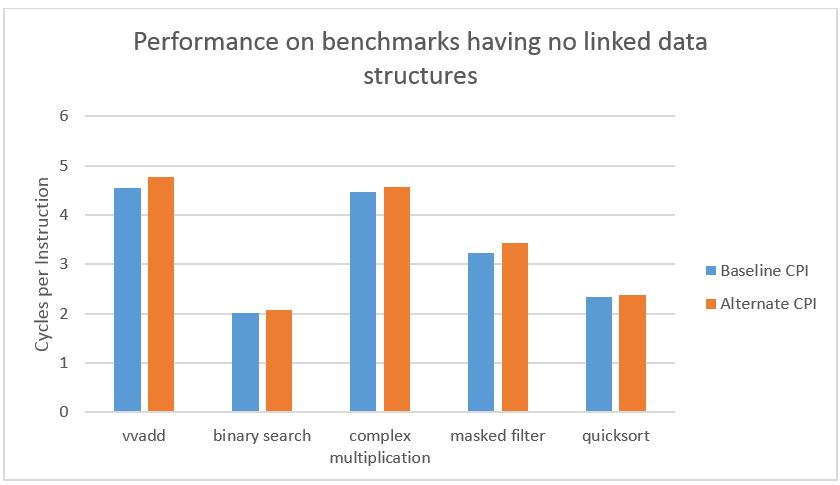}
\caption{Performance comparison between baseline and alternate design for benchmarks having
non-pointer based data structures}
\label{fig:perf-orig}
\end{figure}

\begin{figure}[h]
\centering
\includegraphics[width=\columnwidth]{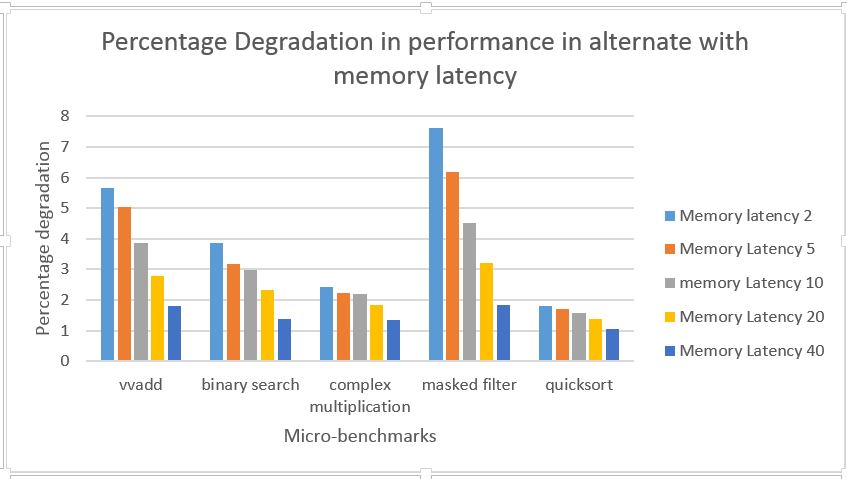}
\caption{ Percentage performance degradation of alternate design over baseline design for non-pointer
based benchmarks}
\label{fig:perf-degrade}
\end{figure}

\begin{figure}[h]
\centering
\includegraphics[width=\columnwidth]{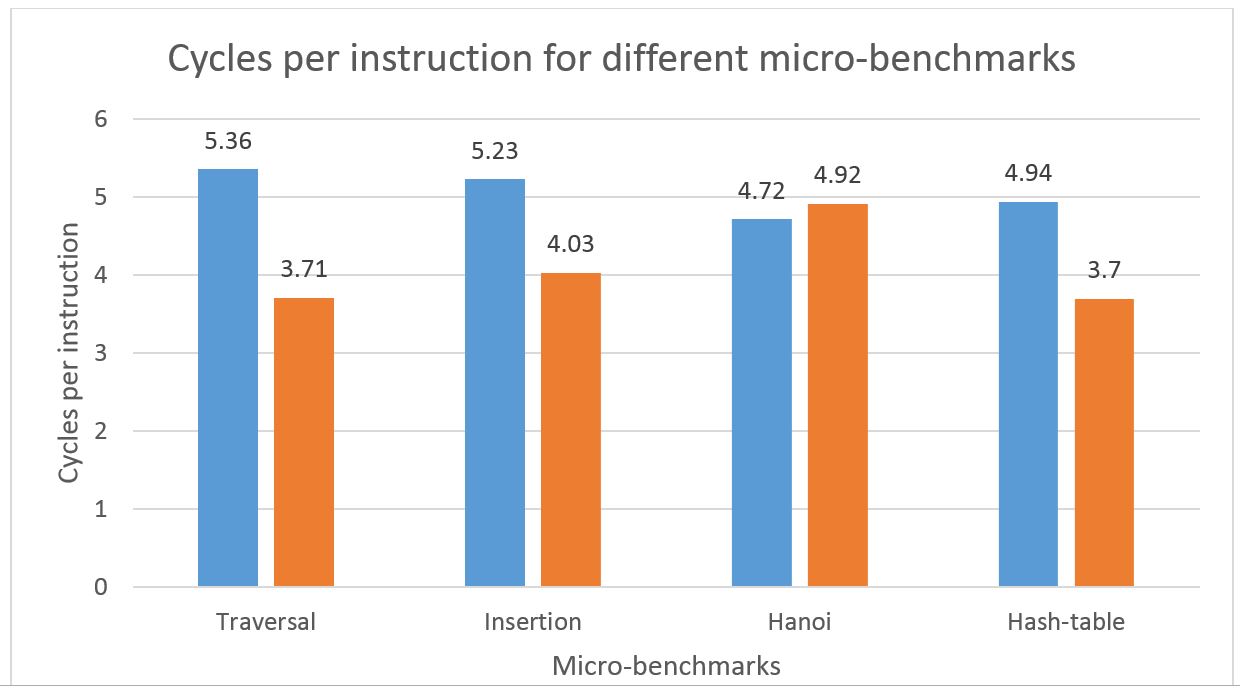}
\caption{Performance comparison for linked-list data structures on Baseline and Alternate design}
\label{fig:bench-perf}
\end{figure}

\begin{figure}[h]
\centering
\includegraphics[width=\columnwidth]{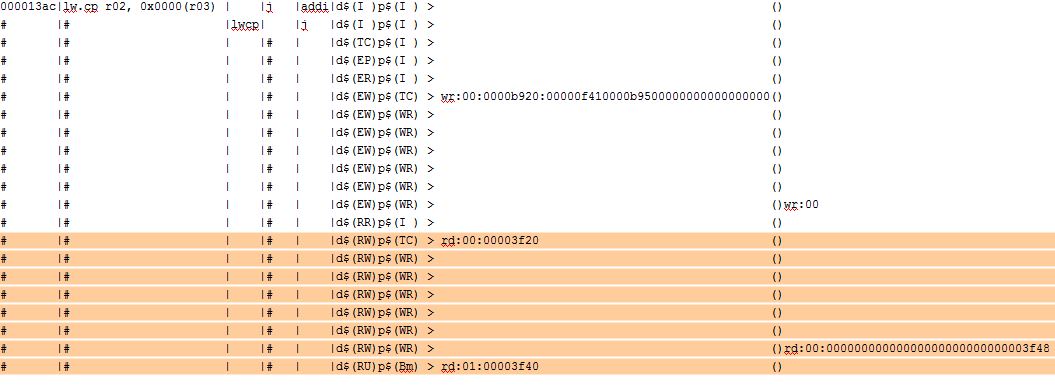}
\caption{ Trace showing first lw.cp miss in the Prefetcher}
\label{fig:first-lwcp}
\end{figure}

\begin{figure}[h]
\centering
\includegraphics[width=\columnwidth]{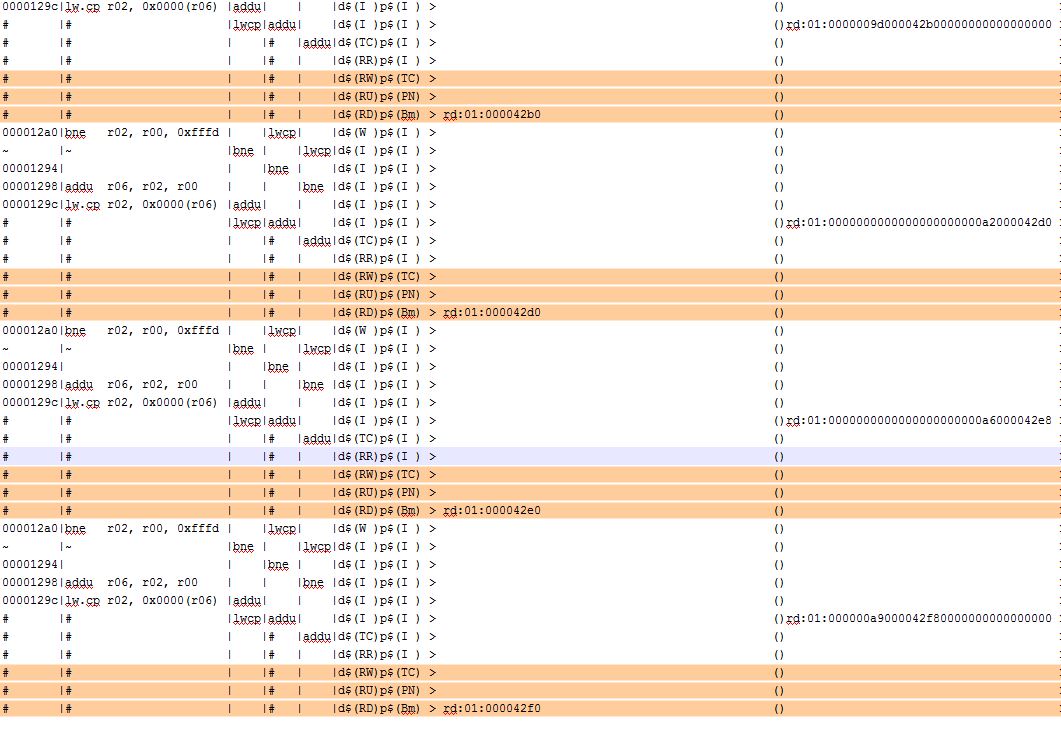}
\caption{Trace showing all consecutive Hits in the Prefetcher}
\label{fig:pfetch-hits}
\end{figure}

To see the impact of the prefetcher on performance, the baseline and the alternate designs were
compared on programs having linked data-structures. For this purpose, linked-list traversal, linked-list
insertion, hash-table and tower-of-hanoi programs were used as benchmarks. As it can be seen from
figure \ref{fig:bench-perf}, the performance improvement was seen for all the benchmarks except the tower-of-hanoi
benchmark. The increase in performance of the alternate design was due to the prefetching of the
next data node in the linked list, as seen from the trace in figure 25 and 26. As seen from the trace in figure \ref{fig:first-lwcp}
the first lw.cp instruction is a miss in both the cache and the prefetcher. However, when the next
node is fetched by the processor it is a miss in the cache but a hit in the prefetcher reducing the
miss penalty of the cache for that node. The highlighted elements in figure \ref{fig:first-lwcp} show the miss penalty
of the prefetcher. In figure \ref{fig:pfetch-hits}, the states of the prefetcher on a read hit are shown. The prefetcher
goes through the three stages highlighted in the trace upon a hit. In the Tag Check (TC) state, the
response is sent back to the cache resulting in the hit latency of the prefetcher of 1 cycle. The cache
then goes to the Refill-Update (RU) and Read-Data Access (RD) states and at the same time the
prefetcher prefetches in the Push-Next (PN) and Buffer to Memory (Bm) states. Thus the prefetcher
can send the next node request to the memory without stalling the cache. This parallel execution
helps in the performance improvement as the prefetching latency gets hidden by the cache latency in
RU and RD states.

The performance degradation for tower-of-hanoi benchmark was observed due to less number of
nodes in the data-structure. The prefetcher helps to reduce the misses which are compulsory and
the ones which are due to eviction of a cache line, when it is replaced (capacity or conflict). For
tower of hanoi benchmark, as the number of nodes are very less, all the nodes are first a miss in
the cache and hit in the prefetcher and the later accesses are hit in the cache. Thus, the overhead
for memory accesses through the prefetcher dominates over the performance improvement due to
prefetching, which leads to poor performance in this case. From figure 36 also, it can be seen that the
number of Read-CP Hits are 6 which is also equal to the number of nodes (disks in the tower) in the
data-structure. Thus the performance benefit of the alternate design over the baseline design depends
a lot on the size of the linked-list data-structure. In most of the real life applications, the linked-list
and tree sizes are quite large and hence the prefetcher may help in the performance improvement in
many of those applications.

\begin{figure}[h]
\centering
\includegraphics[width=\columnwidth]{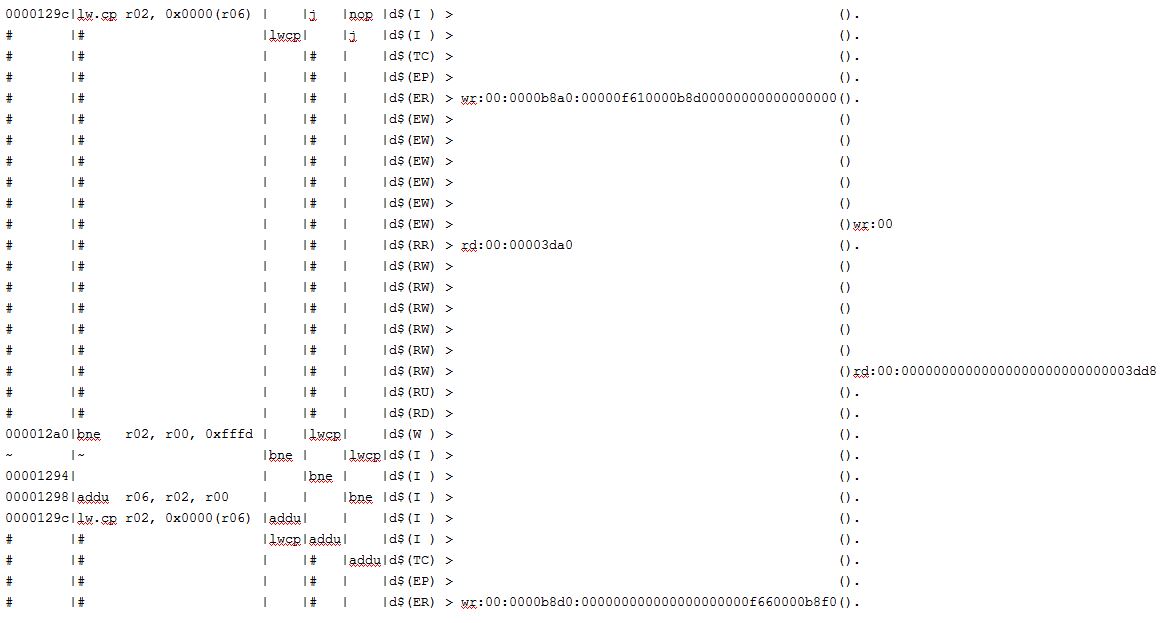}
\caption{Trace for Baseline design}
\label{fig:base-trace1}
\end{figure}

\begin{figure}[h]
\centering
\includegraphics[width=\columnwidth]{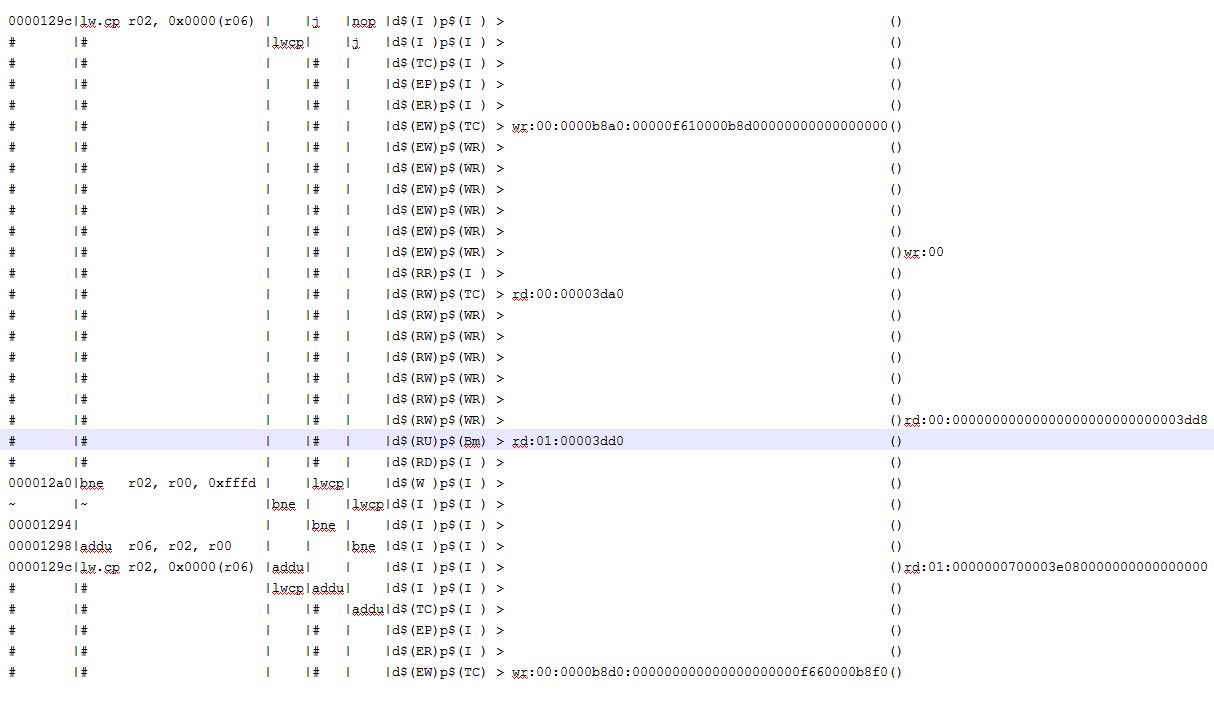}
\caption{Trace for Alternate design}
\label{fig:alt-trace1}
\end{figure}

A trace showing how the memory can be used more efficiently in the alternate design as compared
to the baseline design can be seen in figure \ref{fig:base-trace1} and \ref{fig:alt-trace1}. From the trace it can be observed that the
cycles when the memory is idle are being utilized for prefetching.

\begin{figure}[h]
\centering
\includegraphics[width=\columnwidth]{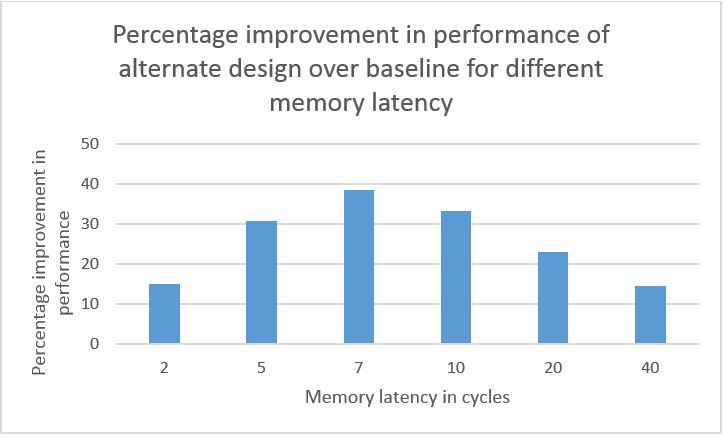}
\caption{Plot showing effect of memory latency on the performance improvement of alternate design
over the baseline design}
\label{fig:mem-lat}
\end{figure}

The memory latency was also varied to study its effect on the performance on linked list based
benchmarks. As the memory latency is increased the effect on the performance becomes more pronounced
as shown in the figure \ref{fig:mem-lat}. This is because while the processor is performing non-memory
operations the prefetcher sends request to the memory for the next node of the linked data-structure
thus hiding the latency of the memory access. This boosts the throughput of the system in these
scenarios. When the memory latency increases too much the performance of the both the system
is severely affected and the prefetcher is unable to completely hide the memory latency with the
non-memory operations running concurrently in the processor. Hence the percentage improvement in
performance is diminished for very high memory latencies.

\begin{figure}[h]
\centering
\includegraphics[width=\columnwidth]{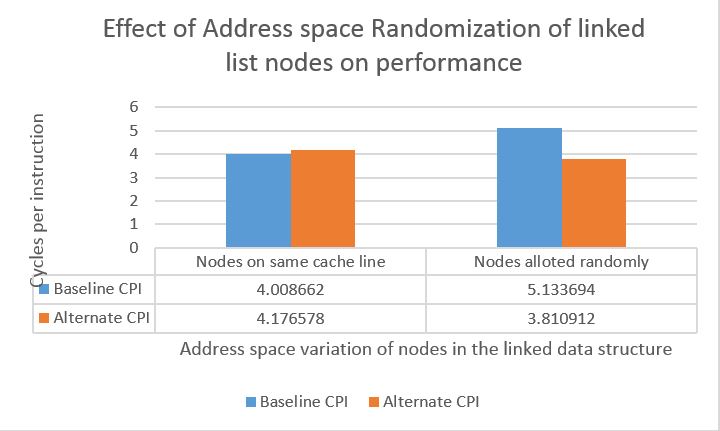}
\caption{Effect of spatial locality in the linked data structure on performance of both the designs}
\label{fig:addr-space}
\end{figure}

Another aspect evaluated was the effect of randomness in the address space of the linked data
nodes. It is observed that the alternate design performs worse as compared to the baseline when two
or more consecutive nodes are on the same cache line. Let us consider the case when two nodes of the
linked list are on the same cache line. When the first read cp request comes from the processor its
a miss in both the cache and the prefetcher. Though the prefetcher fetches the second node, it is
already in the cache when the processor request it. So the extra work done by the prefetcher was not
useful. However, as the spatial locality between the nodes decreases. We see that the performance of
the alternate design increases as shown in figure \ref{fig:addr-space}.

\begin{figure}[h]
\centering
\includegraphics[width=\columnwidth]{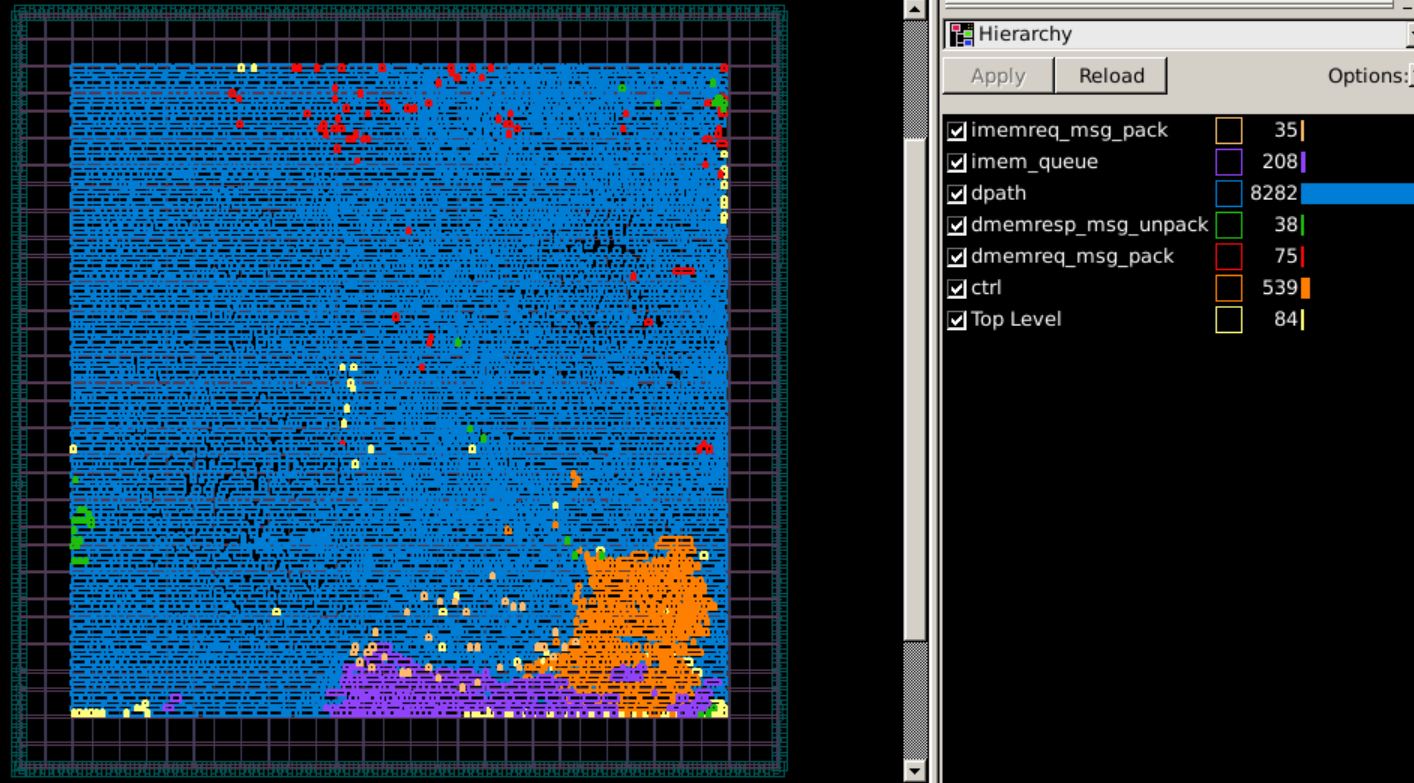}
\caption{Amoeba Plot for Processor}
\label{fig:proc-amoeba}
\end{figure}

\begin{figure}[h]
\centering
\includegraphics[width=\columnwidth]{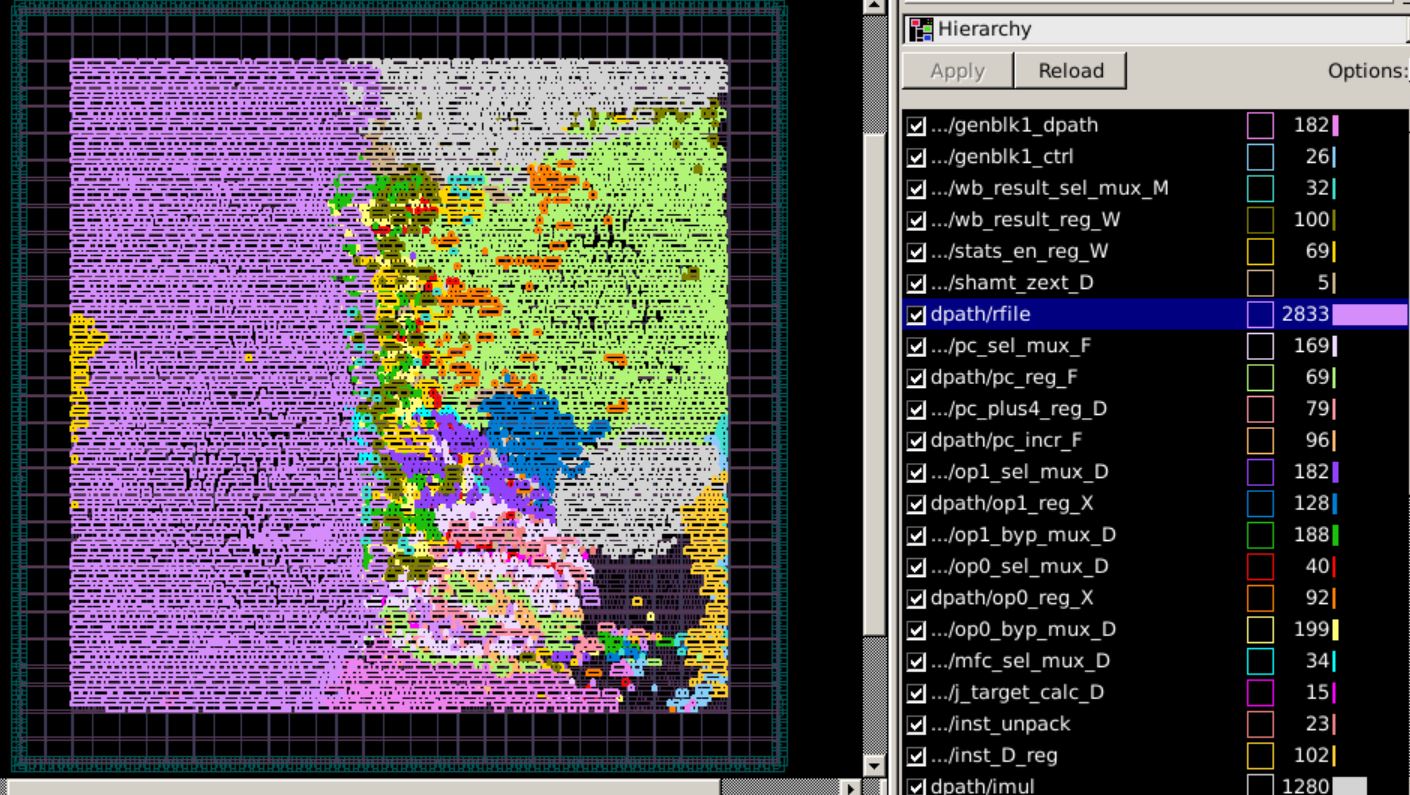}
\caption{Datapath for the Processor showing the Register File}
\label{fig:proc-dpath-amoeba}
\end{figure}

\begin{figure}[h]
\centering
\includegraphics[width=\columnwidth]{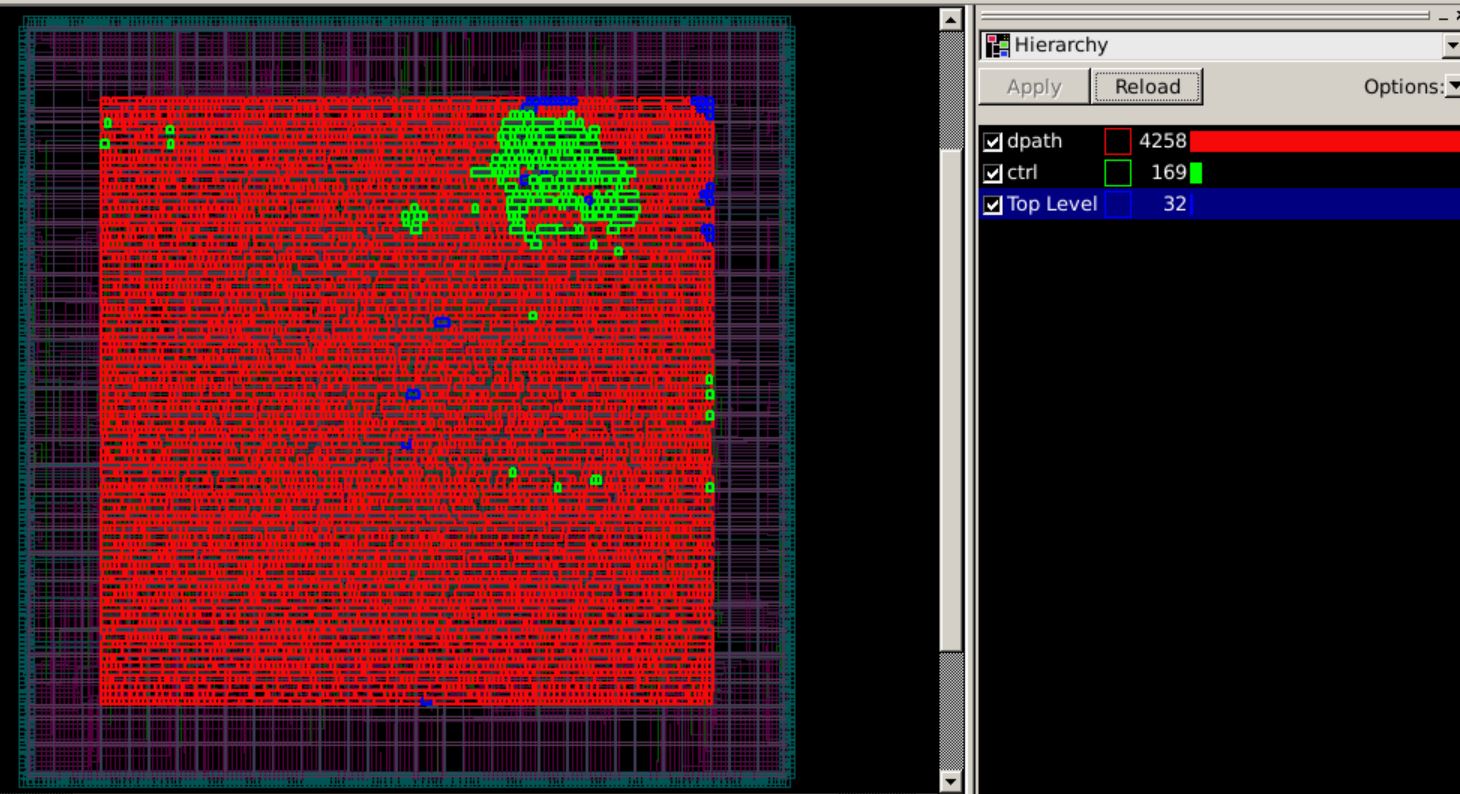}
\caption{Amoeba Plot for Prefetcher}
\label{fig:pfetch-amoeba-all}
\end{figure}

\begin{figure}[h]
\centering
\includegraphics[width=\columnwidth]{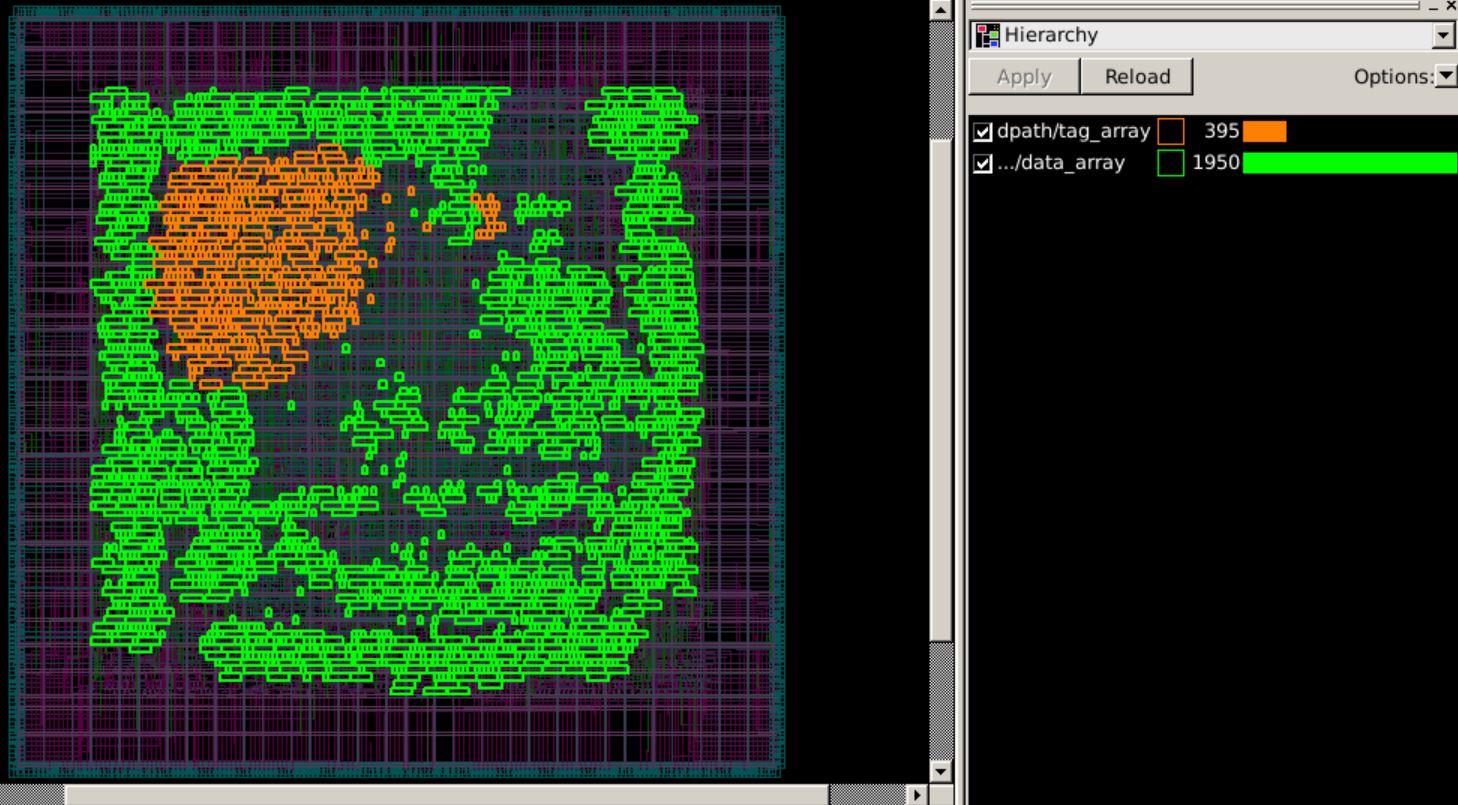}
\caption{Datapath for the Prefetcher showing the Data and Tag arrays}
\label{fig:pfetch-tag-array}
\end{figure}

For area, energy and cycle time analysis, the processor and the prefetcher were pushed
through the ASIC tool flow. For the processor the timing constraint met was 2.0 ns which corresponded
to the area of 114309 $\mu m^2$
. For the prefetcher, it was necessary that its cycle time should not
exceed 2.0 ns so that it does not degrade the performance of the whole processor-cache-prefetchermemory
system. To achieve an optimum area for the prefetcher the timing constraint for the clock
was set to be 2.0 ns i.e. same as that of the processor. The prefetcher design was able to meet the
timing constraint with an area overhead of 51057 $\mu m^2$
.
Figures \ref{fig:proc-amoeba} and \ref{fig:proc-dpath-amoeba} show the amoeba plot and the Datapath for the Processor. It can be seen from
the figure that almost half of the area of the processor is occupied by the Register File. Figure \ref{fig:pfetch-amoeba-all}
and \ref{fig:pfetch-tag-array} show the amoeba plot and data and tag arrays for the prefetcher. Here also almost half of the
area is occupied by the Register files for data and tag arrays. The processor consists of a Register
file of size 32 with 32 bit entries (i.e. total 128 bytes) while the prefetcher consists of a data array of
size 128 bits and 4 entries and tag array of size 26 bits and 4 entries, i.e. total 80 bytes. Hence the
area for the prefetcher was expected to be half of the processor. The results also show that the area
of the prefetcher is 51057 $\mu m^2$ which 44\% of the processor area 114309 $\mu m^2$
. The area for the cache
implementation using SRAMs using CACTI tool was found to be 300507 $\mu m^2$ which is 262\% (around twice) the area of the processor. As the
processor-cache-prefetcher system consists of an instruction and a data cache, the total area of for
our baseline implementation would be around 715323 $\mu m^2$
. The area overhead for the prefetcher is
then 7\% of the total area.

\begin{figure}[h]
\centering
\includegraphics[width=\columnwidth]{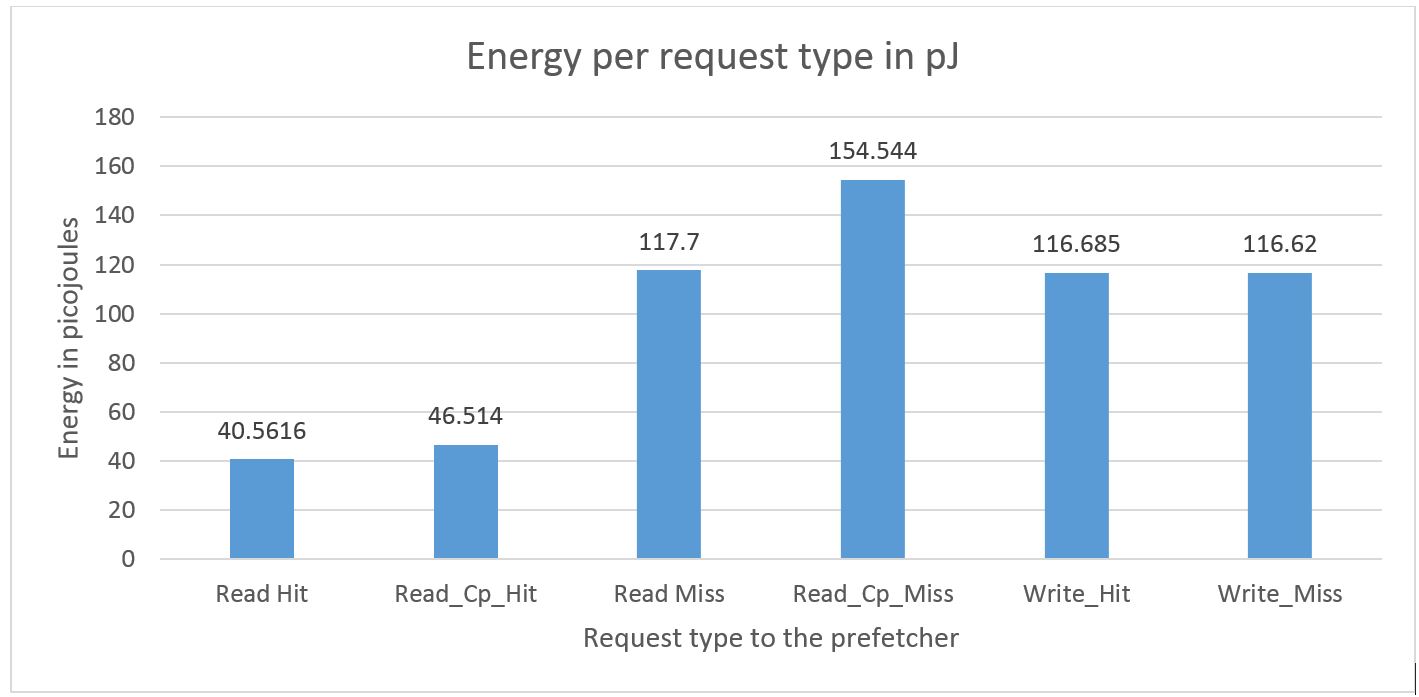}
\caption{Energy consumption of prefetcher for different type of requests}
\label{fig:energy-pj}
\end{figure}

The energy consumed by the prefetcher for different type of requests was obtained (as seen in
the figure \ref{fig:energy-pj}). All the requests to the prefetcher must pass through the Tag Check state, hence all
of them access the data and tag arrays. The energy consumption was seen to be minimum for the
Read Hit request, which was expected, as it only involves the tag and data array access. The energy
consumption for ReadCP Hit came out to be slighlty more than Read Hit due to the extra work done in storing the next address in the buffer and sending prefetch request to the memory. The energy
consumption for Read Miss, Write Hit and Write Miss were found to be almost the same which was
expected as all of them require memory accesses along with data and tag array access. It can also be
seen that the energy consumption for Write Hit is more than Write Miss as the former one requires
invalidating the tag-valid bit for the corresponding address in the tag array which results in extra
work. The energy consumption for Read CP Miss is the maximum which was also expected as it
involves all the possible paths in the datapath like the data and tag array access, request to the
memory and storing the next address in the buffer and sending prefetch request to the memory.

\begin{figure}[h]
\centering
\includegraphics[width=\columnwidth]{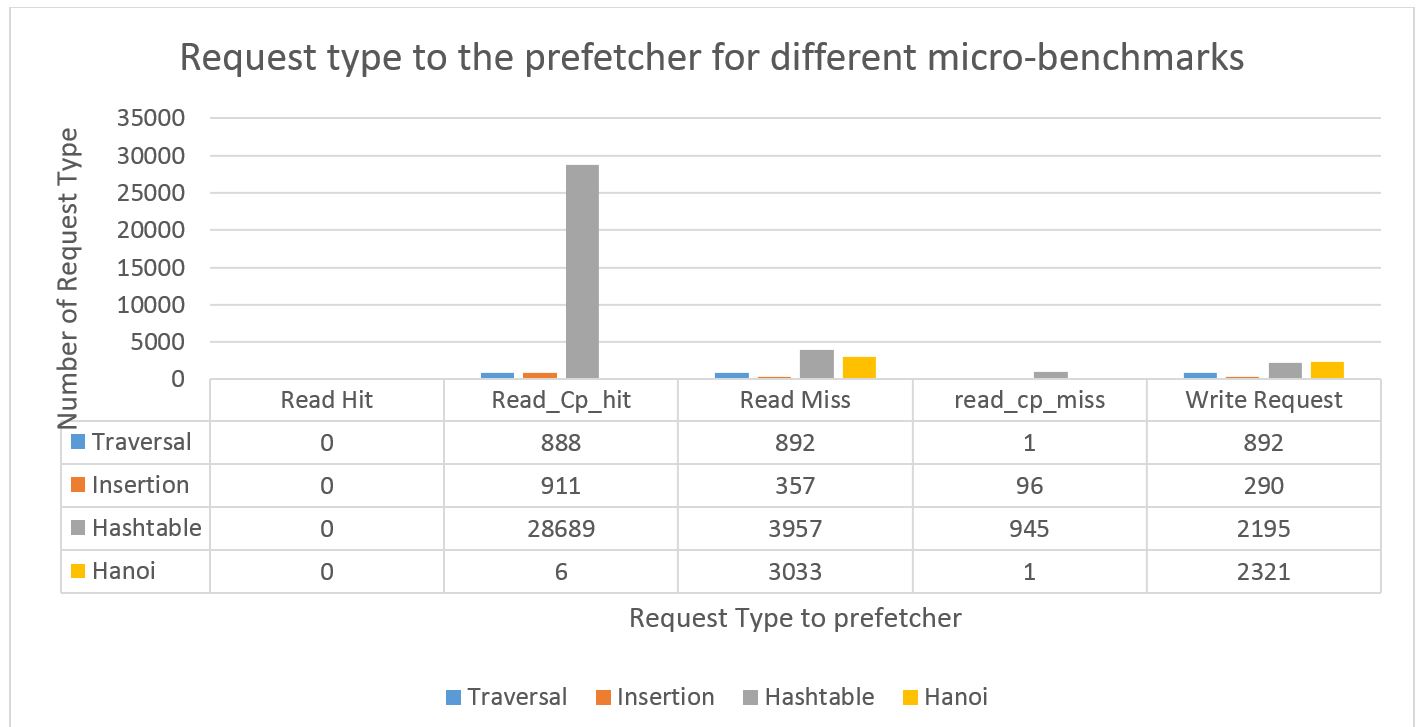}
\caption{Number of different prefetcher requests for different benchmarks}
\label{fig:req-types}
\end{figure}

As the cache was non-synthesizable, running the benchmarks on the entire processor-cache-prefetcher
system to obtain the energy consumption for different benchmarks was not possible. Hence to obtain
the energy consumption of the prefetcher for different benchmarks the exact Read-Write access patterns
along with the information that which one is a hit and which one is a miss had to be obtained.
So, the number of Read Hits, Read CP Hits, Read Misses, Read CP Misses and the Write requests
were extracted from the traces of the benchmarks as shown in figure \ref{fig:req-types}

\begin{figure}[h]
\centering
\includegraphics[width=\columnwidth]{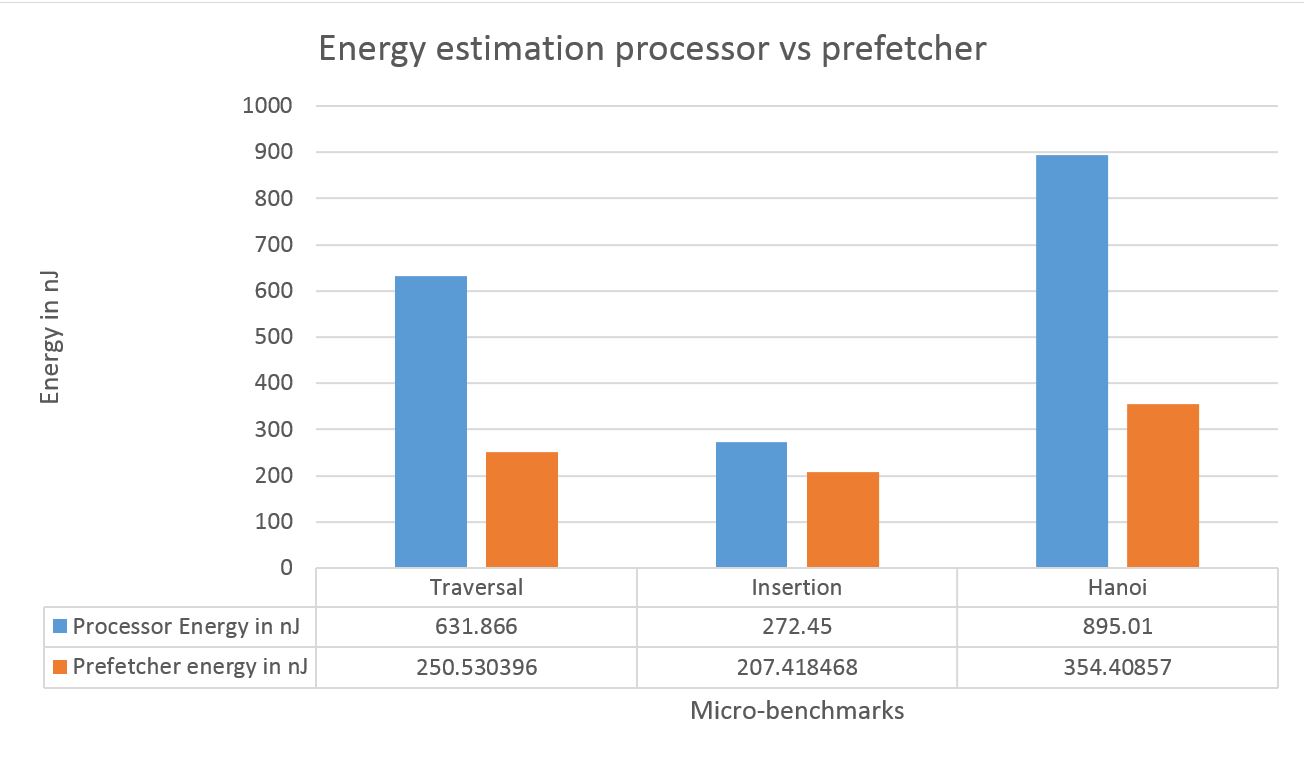}
\caption{ Energy consumption in processor and prefetcher for different benchmarks}
\label{fig:energy-cons}
\end{figure}

Using the number of different type of requests to the prefetcher (figure \ref{fig:req-types}) and the energy consumption
for that particular request from figure \ref{fig:energy-pj}, the energy consumption in the prefetcher was
estimated for different benchmarks. These benchmarks were then ran on the processor in isolation
to obtain the energy consumption in the processor. Figure \ref{fig:energy-cons} shows the energy consumption in
prefetcher and processor for different benchmarks. From the figure \ref{fig:energy-cons}, it can be seen that the energy
consumption from the prefetcher is comparable to the energy consumption from the processor. However
the energy consumption from the cache is expected to be more as compared to the processor
and prefetcher due to its large size and more number of data accesses as compared to the prefetcher.
Since, the cache could not be synthesized we cannot evaluate its energy consumption quantatively.

The critical path of the system was found to be in the processor datapath. The critical path was
found to be the bypass path starting from operand register in execute satge, going through ALU,
execution result select mux, bypass mux in the Decode stage, PC mux select and ending at the instruction
memory request address. This was expected because of the huge bypass muxes and ALU unit.

\section{Conclusion}
\label{sec-conclusions}

From the evaluation, it can be seen that the prefetcher is able to improve the performance of the
system for linked data-structures at the cost of small area and energy overhead. For programs not
including the linked data-structures the performance degradation is quite less. It is also observed
that providing a hint to the hardware from the software can be a very powerful tool in improving the
performance without significantly complicating the hardware.




\begin{thebibliography}{00}


\ifx \showCODEN    \undefined \def \showCODEN     #1{\unskip}     \fi
\ifx \showDOI      \undefined \def \showDOI       #1{#1}\fi
\ifx \showISBNx    \undefined \def \showISBNx     #1{\unskip}     \fi
\ifx \showISBNxiii \undefined \def \showISBNxiii  #1{\unskip}     \fi
\ifx \showISSN     \undefined \def \showISSN      #1{\unskip}     \fi
\ifx \showLCCN     \undefined \def \showLCCN      #1{\unskip}     \fi
\ifx \shownote     \undefined \def \shownote      #1{#1}          \fi
\ifx \showarticletitle \undefined \def \showarticletitle #1{#1}   \fi
\ifx \showURL      \undefined \def \showURL       {\relax}        \fi
\providecommand\bibfield[2]{#2}
\providecommand\bibinfo[2]{#2}
\providecommand\natexlab[1]{#1}
\providecommand\showeprint[2][]{arXiv:#2}

\bibitem[\protect\citeauthoryear{Ebrahimi, Mutlu, and Patt}{Ebrahimi
  et~al\mbox{.}}{2009}]%
        {ebrahimi2009techniques}
\bibfield{author}{\bibinfo{person}{Eiman Ebrahimi}, \bibinfo{person}{Onur
  Mutlu}, {and} \bibinfo{person}{Yale~N Patt}.}
  \bibinfo{year}{2009}\natexlab{}.
\newblock \showarticletitle{Techniques for bandwidth-efficient prefetching of
  linked data structures in hybrid prefetching systems}. In
  \bibinfo{booktitle}{{\em High Performance Computer Architecture, 2009. HPCA
  2009. IEEE 15th International Symposium on}}. IEEE, \bibinfo{pages}{7--17}.
\newblock


\bibitem[\protect\citeauthoryear{Hughes and Adve}{Hughes and Adve}{2001}]%
        {hughes2001memory}
\bibfield{author}{\bibinfo{person}{Christopher~J Hughes} {and}
  \bibinfo{person}{Sarita Adve}.} \bibinfo{year}{2001}\natexlab{}.
\newblock \showarticletitle{Memory-side prefetching for linked data
  structures}.
\newblock \bibinfo{journal}{{\em Urbana\/}}  \bibinfo{volume}{51}
  (\bibinfo{year}{2001}), \bibinfo{pages}{61801--2987}.
\newblock


\bibitem[\protect\citeauthoryear{McKinley, Carr, and Tseng}{McKinley
  et~al\mbox{.}}{1996}]%
        {mckinley1996improving}
\bibfield{author}{\bibinfo{person}{Kathryn~S McKinley}, \bibinfo{person}{Steve
  Carr}, {and} \bibinfo{person}{Chau-Wen Tseng}.}
  \bibinfo{year}{1996}\natexlab{}.
\newblock \showarticletitle{Improving data locality with loop transformations}.
\newblock \bibinfo{journal}{{\em ACM Transactions on Programming Languages and
  Systems (TOPLAS)\/}} \bibinfo{volume}{18}, \bibinfo{number}{4}
  (\bibinfo{year}{1996}), \bibinfo{pages}{424--453}.
\newblock


\bibitem[\protect\citeauthoryear{Roth, Moshovos, and Sohi}{Roth
  et~al\mbox{.}}{1998}]%
        {roth1998dependence}
\bibfield{author}{\bibinfo{person}{Amir Roth}, \bibinfo{person}{Andreas
  Moshovos}, {and} \bibinfo{person}{Gurindar~S Sohi}.}
  \bibinfo{year}{1998}\natexlab{}.
\newblock \showarticletitle{Dependence based prefetching for linked data
  structures}.
\newblock \bibinfo{journal}{{\em ACM SIGOPS Operating Systems Review\/}}
  \bibinfo{volume}{32}, \bibinfo{number}{5} (\bibinfo{year}{1998}),
  \bibinfo{pages}{115--126}.
\newblock


\end{thebibliography}


\end{document}